\documentclass[pra,aps,twocolumn,nopacs,superscriptaddress,nofootinbib]{revtex4-1}
\usepackage{amsmath}  \usepackage{amssymb}  \usepackage{amsfonts}  \usepackage{bm}  \usepackage{bbm}   \usepackage{braket}  \usepackage{color}  \usepackage{comment}  \usepackage{dcolumn}  \usepackage{enumerate}  \usepackage{epsfig}  \usepackage{gensymb}  \usepackage{graphicx}  \usepackage{indentfirst}  \usepackage{lmodern}  \usepackage{mathrsfs}  \usepackage{mathtools}  \usepackage{psfrag}  \usepackage{pst-all}  \usepackage{soul}  \usepackage{units}  \usepackage{xcolor}  \usepackage{ulem}
\usepackage{float} %---APS---SI---arxiv
\usepackage[colorlinks,linkcolor=blue,citecolor=blue,urlcolor=blue,hyperindex,driverfallback=dvipdfm]{hyperref}  \usepackage[T1]{fontenc} %---APS---ACS---SI---arxiv

\def\ii{{\rm i}}  \def\ee{{\rm e}}

        \def\Eb{{\bf E}}                      \def\kb{{\bf k}}    \def\mb{{\bf m}}    \def\pb{{\bf p}}      \def\Rb{{\bf R}}  \def\rb{{\bf r}}       %--- bold vectors
\def\xx{\hat{\bf x}}  \def\yy{\hat{\bf y}}  \def\zz{\hat{\bf z}}  \def\nn{\hat{\bf n}}  \def\rr{\hat{\bf r}}  \def\nn{\hat{\bf n}}  \def\eh{\hat{\bf e}}    
  \def\kparb{{\bf k}_\parallel} %--- unit vectors    
\def\rp{r_{\rm p}}  \def\rs{r_{\rm s}} %--- Fresnel's coefficients
\def\sb{{\bf s}}  

\begin{document}
\title{$\mu$eV electron spectromicroscopy using free-space light
}

% --- affiliations ----------------------------------------

\author{Yves~Auad}
\affiliation{Universit\'e Paris-Saclay, CNRS, Laboratoire de Physique des Solides, 91405, Orsay, France} %---

\author{Eduardo~J.~C.~Dias}
\affiliation{ICFO-Institut de Ciencies Fotoniques, The Barcelona Institute of Science and Technology, 08860 Castelldefels (Barcelona), Spain} %---

\author{Marcel~Tenc\'e}
\affiliation{Universit\'e Paris-Saclay, CNRS, Laboratoire de Physique des Solides, 91405, Orsay, France} %---

\author{Jean-Denis~Blazit}
\affiliation{Universit\'e Paris-Saclay, CNRS, Laboratoire de Physique des Solides, 91405, Orsay, France} %---

\author{Xiaoyan~Li}
\affiliation{Universit\'e Paris-Saclay, CNRS, Laboratoire de Physique des Solides, 91405, Orsay, France} %---

\author{Luiz~Fernando~Zagonel}
\affiliation{Applied Physics Department, “Gleb Wataghin” Institute of Physics, University of
Campinas – UNICAMP, 13083-859 Campinas, SP, Brazil} %---

\author{Odile~St\'ephan}
\affiliation{Universit\'e Paris-Saclay, CNRS, Laboratoire de Physique des Solides, 91405, Orsay, France} %---

\author{Luiz~H.~G.~Tizei}
\affiliation{Universit\'e Paris-Saclay, CNRS, Laboratoire de Physique des Solides, 91405, Orsay, France} %---

\author{F.~Javier~Garc\'{\i}a~de~Abajo}
\email{javier.garciadeabajo@nanophotonics.es} %---
\affiliation{ICFO-Institut de Ciencies Fotoniques, The Barcelona Institute of Science and Technology, 08860 Castelldefels (Barcelona), Spain} %---
\affiliation{ICREA-Instituci\'o Catalana de Recerca i Estudis Avan\c{c}ats, Passeig Llu\'{\i}s Companys 23, 08010 Barcelona, Spain} %---

\author{Mathieu~Kociak}
\email{mathieu.kociak@universite-paris-saclay.fr} %---
\affiliation{Universit\'e Paris-Saclay, CNRS, Laboratoire de Physique des Solides, 91405, Orsay, France} %---

% --- abstract --------------------------------------------
\begin{abstract}
The synergy between free electrons and light has recently been leveraged to reach an impressive degree of simultaneous spatial and spectral resolution, enabling applications in microscopy and quantum optics. However, the required combination of electron optics and light injection into the spectrally narrow modes of arbitrary specimens remains a challenge. Here, we demonstrate microelectronvolt spectral resolution in the nanoscale mapping of photonic modes with quality factors as high as $10^4$. We rely on mode-matching of a tightly focused laser beam to whispering gallery modes to achieve a $10^8$-fold increase in light-electron coupling efficiency. By adapting the shape and size of free-space optical beams to address specific physical questions, our approach allows us to interrogate any type of photonic structure with unprecedented spectral and spatial detail.
\end{abstract}

\maketitle
\date{\today}

% =========================================================
% --- introduction ----------------------------------------
% =========================================================
\section{Introduction}

Thanks to a sustained series of impressive advances in instrumentation \cite{BDK02,KLD14,LTH17,HRK20,LHH21}, electron microscopes can currently focus 60-300\,keV electrons down to sub-{\AA}ngström focal spots with an energy spread of just a few meV. As impressive as this spectral resolution might seem, optical modes of high quality factor $Q$, which are of utmost importance for applications including quantum optics and optical metrology, possess substantially smaller linewidths and, therefore, are unresolvable by state-of-the-art electron spectroscopies such as energy-loss spectroscopy (EELS), cathodoluminescence (CL) \cite{RDF11,GTA13,paper251}, and photon-induced near-field electron microscopy (PINEM) \cite{BFZ09,paper151,PLZ10,FES15,KLS20}. 

The so-called electron energy-gain spectroscopy (EEGS) was proposed \cite{paper114} as a technique that can dramatically enhance electron-based spectroscopies by inheriting the spectral resolution of laser sources while retaining the spatial resolution of electron beams (e-beams). EEGS data thus consist of a series of conventional EELS spectra that are acquired as one scans the wavelength of an external laser irradiating the specimen. Electron-light coupling is mediated by near-field optical components, whose strength is dependent on the optical response of the sample. The latter is consequently retrieved from the intensity associated with light-induced electron energy-gain events as a function of laser wavelength \cite{paper114,paper221}, with an energy resolution that is only limited by the energy-time photon uncertainty ($\sim\hbar\sim1\,$eV\,fs). Early attempts to demonstrate EEGS measurements were performed with 100s-fs laser pulses \cite{paper306,WDS20}, therefore reaching a spectral resolution of tens of meV, yet not overtaking  the few-meV resolution of modern electron monochromators \cite{KLD14}. In parallel, efforts were undertaken to develop EEGS using nanosecond \cite{paper325} or continuous-wave \cite{LWH19} visible laser sources combined with continuous e-beams, although these attempts did not introduce any wavelength tunability. Recently, a spectral resolution of a few $\mu$eV has been demonstrated using on-purpose designed photonic waveguides operating in the near-infrared \cite{HRF21}. By injecting continous laser light, this work elegantly circumvented the fundamental problem of optical coupling to a high-$Q$ cavity. Nevertheless, a sample holder equipped with a dedicated optical fiber was needed to achieve efficient coupling to a photonic device of 10s~$\mu$m in size, thus limiting its applicability to a limited range of specimens.

Here, we demonstrate high-spectral-resolution EEGS enhanced by mode matching between a free-space laser beam and the sample, which renders the technique generally applicable to any kind of specimen. Specifically, we excite whispering-gallery modes (WGMs) of well-defined angular momenta in spherical resonators by means of a focused off-axis laser Gaussian beam using a high-numerical-aperture mirror. We first concentrate on $\sim4~\mu$m silica spheres with quality factors $Q\sim~100-300$, in which EELS and CL characterization reveals sharp resonances \cite{paper383} that are corroborated by EEGS with a higher spectral resolution of $\sim2~$meV. This test system shows that optimum laser-mode coupling is achieved via conservation of angular momentum, resulting in a $10^8$-fold enhancement of the coupling efficiency relative to irradiation by an unfocused light plane wave. The improved sensitivity of EEGS is also explained in terms of the sharp laser linewidth ($7~\mu$eV) and high numerical aperture of the light injection system. We then demonstrate the full potential of this approach by controlling the laser beam position with sub-$\mu$m accuracy and resolving narrow optical modes ($Q\sim10^4$) in EEGS spectra of $\sim8~\mu$m polystyrene spheres, which are unobservable by EELS or CL. Our technique is readily applicable to study arbitrary structures and represents an increase by more than two orders of magnitude in spectral resolution relative to state-of-the-art EELS with the same spatial resolution.

% Figure 1 ------------------------------------------------
\begin{figure*}
    \centering
    \includegraphics[width=0.65\textwidth]{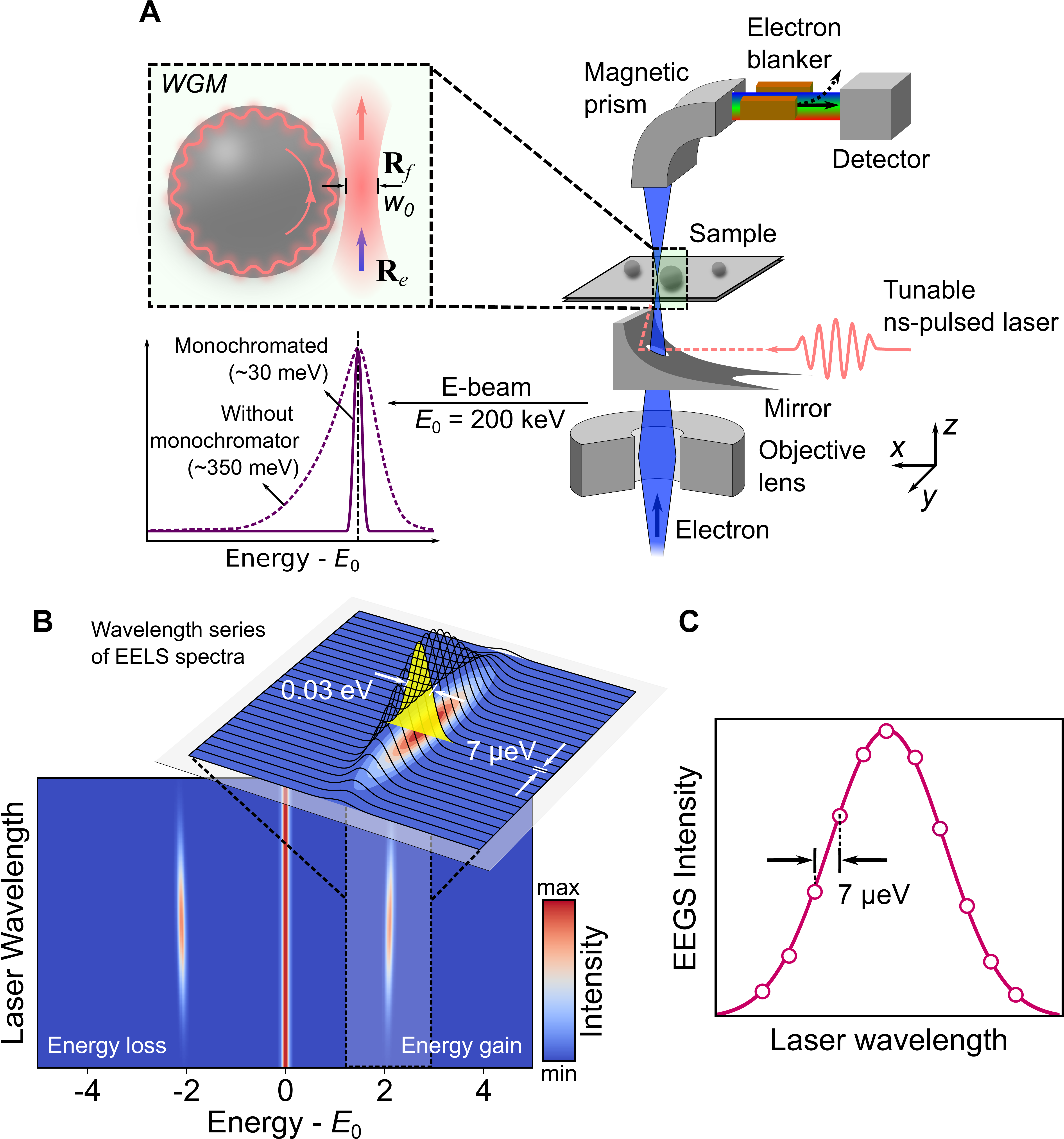}
    \caption{\textbf{Versatile EEGS experiments using a continuous electron gun microscope.} ({\bf A}) Schematics of the setup. A laser beam is focused down to a $\sim1~\mu$m spot and positioned with sub-$\mu$m precision at $\Rb_f$ on the sample of interest by using a parabolic mirror. A monochromatized 200~keV electron is also focused on the sample and scanned to acquire EELS, EEGS, and CL signals. EEGS measurements are taken by synchronizing light (through a laser trigger) and electrons (through an e-beam blanker) at the detector. ({\bf B}) A series of EELS spectra is acquired for a given electron probe position as the laser wavelength is scanned over the spectral region of interest. The strength of the energy-gain signal is boosted when the light wavelength is close to an optical resonance of the specimen. ({\bf C}) Using a wavelength-independent light power, the energy-gain signal yields the EEGS intensity as a function of light wavelength.} \label{fig:principle}
\end{figure*}

% Figure 2 ------------------------------------------------
\begin{figure*}
    \centering
    \includegraphics[width=0.65\textwidth]{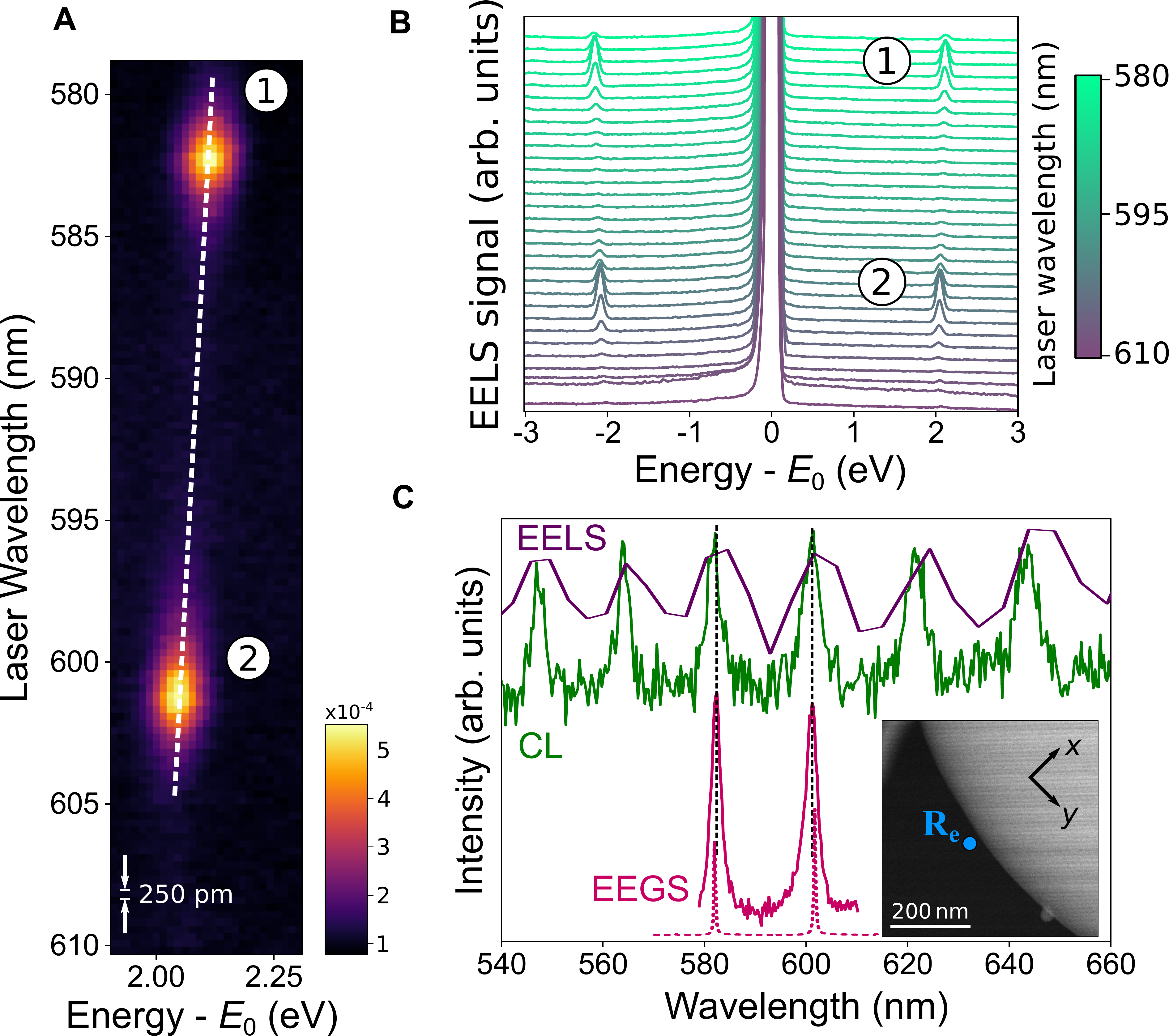}
    \caption{\textbf{EEGS measurements in WGM resonators with $Q\sim$ 100--300}. ({\bf A}) Measured series of EELS gain spectra as a function of light wavelength (vertical axis), showing two distinct resonances at 586.27~nm (2.1296~eV) and 601.02~nm (2.0631~eV) with quality factors of 244 and 194, respectively. ({\bf B}) Cascade of one in every four spectra taken from (A), where both the stimulated loss and gain sidebands are observable. ({\bf C}) EEGS, EELS, and CL measured spectra (solid curves) compared to theory (dashed curves) for the same e-beam probe position (blue circle at $\Rb_e$ in the inset) and mirror focal position.
} \label{fig:mediumQ}
\end{figure*}

% Figure 3 ------------------------------------------------
\begin{figure*}
    \centering
    \includegraphics[width=0.6\textwidth]{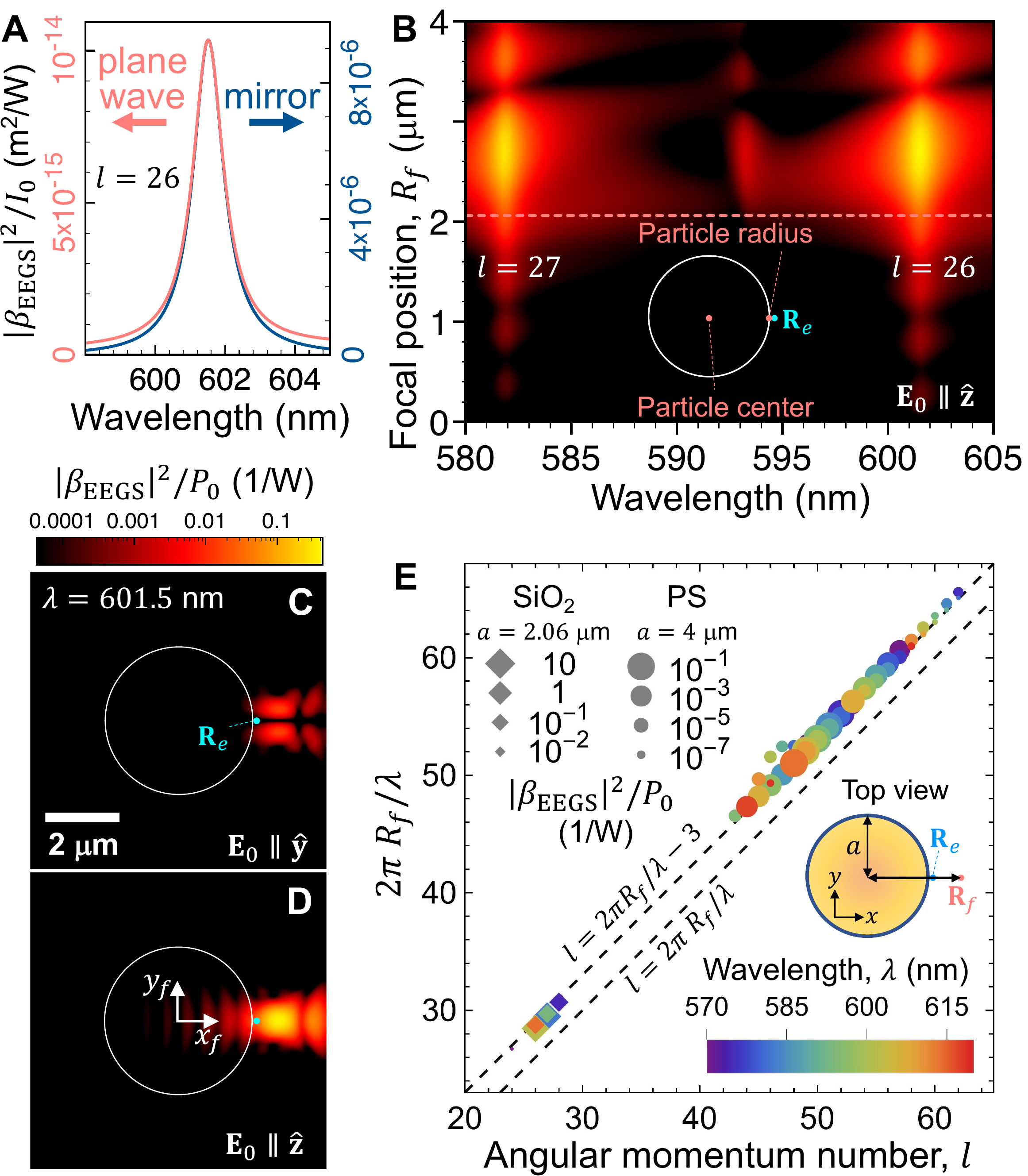}
    \caption{\textbf{Free-space optical mode matching}. Dependence of the EEGS probability on the illumination conditions for a SiO$_2$ sphere of 4122~nm diameter.
    ({\bf A}) EEGS probability around a 601.5~nm resonance for plane-wave and mirror-focused illumination, normalized to the light intensity incident on the sphere and mirror, respectively.
    ({\bf B}) Dependence of the EEGS probability on light wavelength and focal spot position. The vertical axis corresponds to the distance from the focal spot to the sphere center as it moves along the $x$ axis (see inset).
    ({\bf C} and {\bf D}) EEGS probability map at the peak wavelength of (A) with ${\bf R}_f=(x_f,y_f)$ scanned for incident light polarized along $y$ and $z$, respectively. The sphere contour is shown as a white circle. The color scale is shared by panels (B) to (D).
    ({\bf E}) Optimum optical-focus position $R_f$ as a function of orbital momentum number $l$ for resonances in PS (solid circles) and SiO$_2$ (open circles) spheres of diameters 8000~nm and 4122~nm, respectively. Symbols show all modes of high-quality factor within the $\lambda=570-620~$nm spectral region (see color-coordinated scale). The EEGS probability is indicated by the symbol size (see legend). The dashed straight lines correspond to $2\pi R_f/\lambda$ equal to $l$ and $l+3$. In (A), (B), and (E), the e-beam passes at a fixed position $\sim80~$nm away from the sphere surface on the $x$ axis, as indicated in the insets. The intensity $I_0$ and power $P_0$ of the light incident on the mirror are related through $P_0=A\,I_0$, where $A=18.74~$mm$^2$ is the mirror area.
} \label{fig:theory}
\end{figure*}

% Figure 4 ------------------------------------------------
\begin{figure*}
    \centering
    \includegraphics[width=0.8\textwidth]{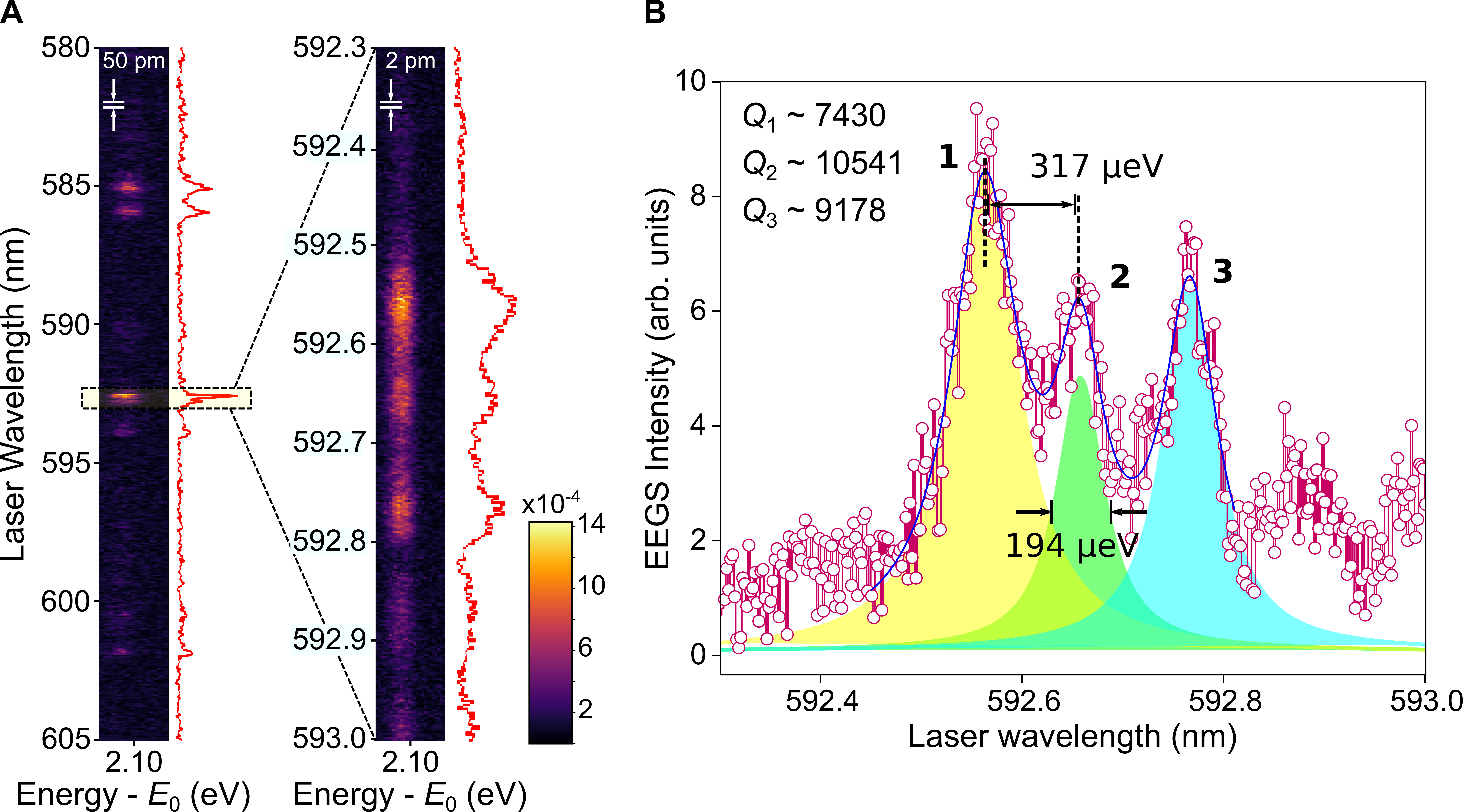}
    \caption{\textbf{High quality factor in large polystyrene spheres}. ({\bf A}) Two laser wavelength series of EELS spectra, taken at the same electron probe position over an increasingly small wavelength range with step sizes of 50~pm and 2~pm, respectively. ({\bf B}) EEGS spectrum constructed from the rightmost EELS series in (A), along with Lorentzian fittings revealing three resonances (1--3) of quality factors 7430, 10541, and 9178, respectively. The uncertainties for the measured quality factor are $<0.1$ for the three resonances.}
    \label{fig:HighQ}
\end{figure*}

% =========================================================
% --- results ---------------------------------------------
% =========================================================
\section{RESULTS AND DISCUSSION}

A sketch of the experimental setup is displayed in Figure~\ref{fig:principle}. Experiments were carried out in a modified Nion Hermes 200 transmission electron microscope (Chromatem) working at 200~keV with a subnanometer e-beam. A nanosecond-laser beam was focused on the sample and synchronized with the EELS detection in order to record PINEM spectra, as detailed in the Appendix. Critical improvements were implemented relative to our previous work \cite{paper325}. Specifically, the microscope was fitted with a high-numerical-aperture (NA$\sim0.5$) Attolight Mönch light detection/injection system able to focus down to a 1~$\mu$m spot size with sub-$\mu$m accuracy. The time-averaged laser input power of $1-5~$mW used in experiments resulted in a typical $10^{9-10}~$W/m$^2$ optical spot intensity. We used the mirror to position the laser spot at the edge of the WGM resonators. In addition, the laser wavelength was tuned to spectrally map the resonances, with a wavelength resolution of 2~pm ($\sim7~\mu$eV at 585~nm), limited by the laser specifications. Because the EEGS signal was weak ($\sim10^{-4}$ of the measured ZLP), a large increase of signal-to-noise ratio was needed, which we achieved through a slight monochromation of $30-50~$meV over the 350~meV initial e-beam energy spread. This led to a strong suppression of the ZLP tails, which would otherwise produce a substantial background. We recorded spectra with a direct electron detector (MerlinEM, from Quantum Detectors) using an effective current of $\sim0.2~$fA (see Appendix), comparable to PINEM experiments. The WGM resonators were drop-cast on a lacey-carbon sample grid, which was coated with 60~nm of silver to improve charge and thermal dissipation. Finally, alignment of the laser spot to the microscope optical axis was achieved with $\sim1~\mu$m precision by maximizing the electron EEGS signal from a featureless silver film. A more detailed description of the setup is offered in the Appendix.

The EEGS electron-light coupling is described by a single parameter \cite{paper371}:
\begin{align}
\beta({\bf R}_e,\omega)=\frac{e}{\hbar \omega} \int_{-\infty}^{\infty}dz\,E_{z}({\bf R}_e,z,\omega)\,e^{-i\omega z/v},
\label{eq:EEGS}
\end{align}
where ${\bf R}_e$ is the transverse electron probe position, $v$ is the electron velocity, $\omega$ is the angular frequency of the external light, and $E_z$ is the optical electric field component along the e-beam direction $z$, corresponding to a time-varying field $E_z({\Rb}_e,t)=2{\rm Re}\{E_{z}({\bf R}_e,z,\omega)e^{-i\omega t}\}$, which is dependent on the focal beam profile and position relative to the specimen (see Appendix). In our experiment, we use a low peak-intensity illumination, such that $|\beta|^2\ll1$ is the probability for the electron to gain one photon quantum (i.e., the EEGS signal is essentially a perturbation).

To illustrate and validate the principle of EEGS and its relation to other spectroscopies, we start by studying $\sim4~\mu$m silica spheres (Figure~\ref{fig:mediumQ}), which are known to exhibit quality factors $Q\sim10^2$ under similar experimental conditions and good electron-WGM coupling for the employed 200~keV electrons \cite{paper383}. In Figure~\ref{fig:mediumQ}A, we plot a measured series of spectra acquired for varying light wavelengths (vertical axis, in steps of 250~pm, corresponding to a 0.92~meV photon-energy interval at 580~nm) with a constant laser power of $\sim1.5~$mW and the e-beam probe positioned as indicated by the blue circle ${\bf R}_e$ in Figure~\ref{fig:mediumQ}C. We observe two distinct WGMs with $Q=244$ and $Q=194$ separated by a spectral distance of 66.4~meV. Due to the high monochromaticity of the e-beam, it is possible to resolve the energy-gain resonance shifting in energy as we raster the laser wavelength (dashed white line in Figure~\ref{fig:mediumQ}A). One of every four of these spectra is shown in the cascade in Figure~\ref{fig:mediumQ}B, where we note the presence of both stimulated electron energy gain and loss features. The gain signal accounts for a fraction $\sim5\times10^{-4}$ of the integrated measured spectrum (see color bar in Figure~\ref{fig:mediumQ}A), implying a gain probability of $\sim5\times10^{-3}$ after correcting for the e-beam blanker time window ($\sim10$ times the laser pulse duration). Furthermore, there are no visible higher-order replicas at multiples of the photon energy $\pm n\hbar\omega$, as expected in the low-intensity regime \cite{paper151,PLZ10}. In Figure~\ref{fig:mediumQ}C, we display the measured EEGS intensity (solid red curve) obtained by integrating the EELS spectra series between 1.9~eV and 2.3~eV for each light wavelength. A similar result is obtained by integrating the stimulated loss peak, although the gain side is free from loss features and thus has a better signal-to-noise ratio. In this particular sample, the $\sim30~$meV spectral resolution of our setup is already enough to unveil the same gallery-modes in EELS, and additionally, modes can be identified by collecting light leakage from the resonator to the far field to record a CL spectrum. The three spectroscopies (EEGS, EELS, and CL) are performed for the same e-beam probe position ${\bf R}_e$ and mirror focal spot ${\bf R}_f$. As expected from the reciprocity of Maxwell's equations, the CL and EEGS spectral variations look similar \cite{paper371} (see Appendix). Also, in non-dissipative systems, the EELS and CL probabilities should be identical because radiation losses are the only source of energy losses \cite{paper149}, although deviations between the two of them can arise because we are collecting only a fraction of the emission solid angles \cite{SLH21}. Nevertheless, no spectral shifts can be discerned between EELS (the equivalent of optical extinction \cite{paper251}), CL (scattering), and EEGS within the single-pixel uncertainty of the EELS channels. Theoretical modelling of the EEGS intensity (Figure~\ref{fig:mediumQ}C, dashed curves; see details in Appendix) matches the WGM positions for a fitted sphere diameter of 4122~nm (consistent with the experimentally determined diameter), although the predicted quality factors are $\sim 1500$, almost one order of magnitude higher than the experimental results, presumably because of the effect of losses produced at the supporting carbon structure. The reported EEGS features with a full width at half maximum (FWHM) of $\sim2~$meV separated by 66.4~meV already demonstrate a spectral resolution one order of magnitude better than the used EELS resolution at 200~keV electron energy, but also better than the ultimate spectral resolution of the machine ($\sim5~$meV at 60~keV).

It should be noted that CL yields a faint signal, which we accumulate for $\sim30~$s without electron monochromation (i.e., using 10--20 times more current than in the monochromated experiments). Although this is faster than EEGS acquisition, the high electron currents used can produce larger sample damage. In addition, the EEGS signal can be enhanced by increasing the incident laser power, which is still well below the sample damage threshold.

We next interrogate the potential of EEGS for the investigation of high-Q photonic modes (i.e., those in which a high spectral resolution is actually required). These modes are weakly coupled to the far field, and therefore, a clear understanding of how to maximize light-electron coupling is needed. We start by presenting theoretical calculations that illustrate the benefits of using a position-controlled high-numerical-aperture focusing system.

The EEGS simulations presented in Figure~\ref{fig:theory}A for the SiO$_2$ sphere studied in Figure~\ref{fig:mediumQ} show an enhancement in probability by eight orders of magnitude when moving from plane wave illumination to focused illumination (assuming the same power and initial laser beam extension over the mirror area in both scenarios). Besides a clear improvement due to the focusing effect of the mirror, the principle behind such huge increase in coupling efficiency can be found in mode-matching between a free-space optical beam and WGMs when the former is focused near the edge of a dielectric sphere \cite{ZDG21}. Such a mode-matching can be well-understood in terms of a preferential orbital angular number $l\sim 2\pi R_f/\lambda$ produced when the focal spot is at a distance $R_f$ from the sphere center and $\lambda$ is the light wavelength. Near the edge of the silica sphere discussed in Figure~\ref{fig:mediumQ}, we have $R_f = 2060 \pm 500$ nm, leading to preferential coupling to modes of angular momentum number $l = 22 \pm 5$, in agreement with the angular order $l = 26$ deduced from theory (see Appendix). The beneficial effects of mode-matching can be further evidenced in several ways. The EEGS signal for an electron passing near the edge of  a dielectric sphere is represented in Figure~\ref{fig:theory}B as a function of the optical beam position for a fixed wavelength, revealing that the signal is strikingly peaked at the position of optimum mode-matching for the value of $l$ corresponding to the  mode that is resonant at a wavelength $\lambda$. In Figure~\ref{fig:theory}(C\&D), the mirror position is scanned with a fixed electron probe position $\Rb_e$ for the two possible light polarizations, again evidencing the mode-matching condition through an optimal mirror position $\Rb_f$. These results are further corroborated by examining multiple resonances comprised in the 570--620~nm range both for the already discussed $4.122~\mu$m SiO$_2$ sphere and for a bigger $8~\mu$m polystyrene (PS) sphere. The relation between the focal point and the angular momentum number is linear as expected \cite{ZDG21}, although shifted by 3, presumably as a result of the finite beam size. More generally, mode matching between focused light and WGM resonators is known to be similar to light coupling to a waveguide \cite{ZSS13_2} and reach coupling efficiencies up to 20$\%$. We thus anticipate that high-$Q$ cavities could be studied with nanometer-scale resolution in a way similar to waveguides \cite{HRF21}, but with higher flexibility regarding the type of specimen and arrangement.

To validate this hypothesis, we examined a larger PS sphere of $8~\mu$m in diameter in search for high-$Q$ resonances. In the one considered in Figure~\ref{fig:HighQ}, EEGS measurements with a laser power of $\sim1~$mW revealed quality factors as high as $10^4$ (FWHM of $194~\mu$eV). In the leftmost wavelength series of Figure~\ref{fig:HighQ}A, the laser step was fixed at 50~pm, scanning a relatively broad energy range $\sim88.34~$meV from 580~nm to 605~nm. After identifying a sharp resonance close to 592.6~nm, the energy range and the laser step were reduced to the limit of our laser, rastering a wavelength range of 0.7~nm in steps of 2~pm (i.e., a spectral step of approximately $7~\mu $eV). The yellow dashed rectangle illustrates the energy range used for the sequential acquisitions. This series of measurements demonstrates the potential of EEGS to map a significant range of resonance linewidths by adapting the laser wavelength scan range and the spectral step. Interestingly, the acquisition of each wavelength series took $\sim8~$min, showing remarkable repeatability of the experiments, with minor changes between them. The rightmost series of Figure~\ref{fig:HighQ}A is integrated along the electron energy axis for each laser wavelength to produce the EEGS intensity curve shown in Figure~\ref{fig:HighQ}B, as well as the Lorentzian fittings for the three sharpest resonances, yielding quality factors of 7430, 10541, and 9178. The peak separation is $\sim300~\mu$eV apart, and the highest-$Q$ mode has a FWHM of $194~\mu$eV.

We remark that EELS cannot resolve such fine features due to its limited spectral resolution. Also, although CL could in principle be performed with a sufficiently accurate light spectrometer, it cannot resolve high-$Q$ features in practice because of its smaller signal count rate. Indeed, the ratio of integrated CL and EEGS probabilities scales as $\Gamma_{\rm CL}/\Gamma_{\rm EEGS}\sim1/Q$ when measuring a mode of quality factor $Q$ (see detailed derivation in the Appendix). In this work, the illumination intensity is $\sim10^8~$W/m$^2$ and the resonance energy $\sim2~$eV, and hence, we have $\Gamma_{\rm CL}/\Gamma_{\rm EEGS}=2.5$ for $Q=200$, thus explaining why CL can resolve the WGMs in the smaller silica spheres discussed in Figure~\ref{fig:mediumQ}. In contrast, $\Gamma_{\rm CL}/\Gamma_{\rm EEGS}=0.05$ for $Q=10^4$, therefore yielding an undetectable CL signal in the larger spheres. This constitutes a compelling argument supporting the superior signal-to-noise ratio of EEGS relative to CL. It is worth mentioning that the same arguments limit the sensitivity of EEGS when it is performed with a laser of small spectral resolution compared to the WGM linewidth (e.g., when using femtosecond light pulses, in which most of the injected photons lie outside the resonance, thus resulting in a dramatic loss of coupling efficiency).

% =========================================================
% --- conclusions -----------------------------------------
% =========================================================
\section{CONCLUSIONS}

The present work demonstrates EEGS with nanometer spatial resolution and down to $200~\mu$eV spectral resolution on arbitrary optical dielectric cavities, therefore leveraging the spatial resolution of free electrons, the versatility of electron microscopy, and the spectral resolution of laser light sources. Our results are made possible by using a small laser bandwidth compared with the widths of the probed modes, as well as by adapting the symmetry, size, and shape of the laser beam to that of the excitations in those cavities, all in a free-space configuration. More general strategies for laser-to-cavity mode matching could rely on light beams sculpted in amplitude and phase by slide projection (e.g., through spatial light modulators). This versatility holds potential for imaging at the ultimate limits of resolution, as required for the characterization of vanguard photonic structures. It should be noted that EEGS is not limited to the visible spectrum, but it should be useful to probe mid- \cite{KDS21} and far-infrared modes, going well beyond the spectral resolution reached by electron monochromation, which is now reaching the limits prophesied by their creators \cite{KUB09}. Applications in vibration mapping at such high resolution should directly impact biological applications \cite{RAM16}. Fast beam blanking technologies are also becoming increasingly available in distinct operation frequencies and duty cycles \cite{VVK18,ZHC19}, enabling this experiment to be implemented in different microscope configurations. In addition, beyond the current design, energy-gain experiments using pulsed laser sources could be performed without e-beam blankers by relying on time-resolved electron detectors \cite{AWB22}. The use of nanosecond-based time-resolved detectors could further increase the spectroscopic portfolio in electron microscopy by means of temporal correlation between photons and electrons, as recently demonstrated in experiment \cite{VAW22,FHA22}. The integration of such new techniques holds promise for a thriving future in free-electron-based nanooptics.

% =========================================================
% --- methods/appendix ------------------------------------
% =========================================================
\begin{widetext} %---arxiv
\appendix

% =========================================================
% --- EEGS experiments ------------------------------------
% =========================================================
\section{Experimental setup for continuous-wave EEGS spectroscopy}

% --- light detection -------------------------------------
\subsection{Light detection}

For the CL experiments, light detection was performed with an Attolight Mönch system fitted with a $\approx$ 0.5 numerical aperture. The light collected through the mirror was focused into a single-step-index multimode fiber of 100~$\mu$m in diameter (model FG105LVA Thorlabs Inc.) coupled to an optical spectrometer of similar numerical aperture, roughly 0.1. The fiber transmission efficiency was more than 95\% within the entire spectral range used throughout this work.

% --- laser and injection ---------------------------------
\subsection{Laser source and laser injection parameters}

For light injection, a dye laser beam (Pyromethane 597 organic solution, with peak efficiency at 585 nm) was focused on the sample by means of a high-numerical-aperture parabolic reflector after being spatially filtered by a single-mode optical fiber. The focused light beam, of approximately $\sim$ 1~$\mu$m in diameter (see Sec.~\ref{parabolicmirror} for a numerical analysis), could be precisely positioned at the edge of a whispering-gallery mode resonator (WGMR; a large dielectric sphere) thanks to a M\"onch three-dimensional stage with an ultimate 50\,nm precision. The pulsed dye laser operated at a repetition rate of 10\,kHz and delivered pulses with a duration in the 20--30\,ns range. In contrast to conventional PINEM experiments in ultrafast electron microscopes, the electron and photon temporal synchronization was carried out by means of a fast blanker placed immediately before the detector, as shown in Fig.~1 in the main text and discussed elsewhere in more detail \cite{paper325}. The nanosecond-resolved beam blanker selected electrons that crossed the sample during the laser pulse duration, with the beam blanker time window fixed at 250 ns in order to accommodate a slight variation of the laser pulse delay on the pump laser current. Notice that only a tenth of the transmitted electrons were exposed to the laser action. Typical time-averaged input powers were in the range of 1--5\,mW, thus resulting in 10$^{8-9}$\,W/m$^2$ light intensities. The acquisition time per EELS spectrum in the wavelength series was typically 100\,ms. Over this time interval, a set of 5--25 measurements were performed (and subsequently averaged) per wavelength. A direct electron detector (Merlin, from Quantum Detectors) was used, which improved the quantum efficiency and the signal-to-noise ratio of the acquired data. The laser pulse duty cycle (i.e., the product of its pulse length and its repetition rate) was $2\times10^{-4}$, implying that the average current available for generating an EEGS signal was typically $2\times10^{-16}$\,A.

% Figure Sexp ---------------------------------------------
\begin{figure}
\centering{\includegraphics[width=0.6\linewidth]{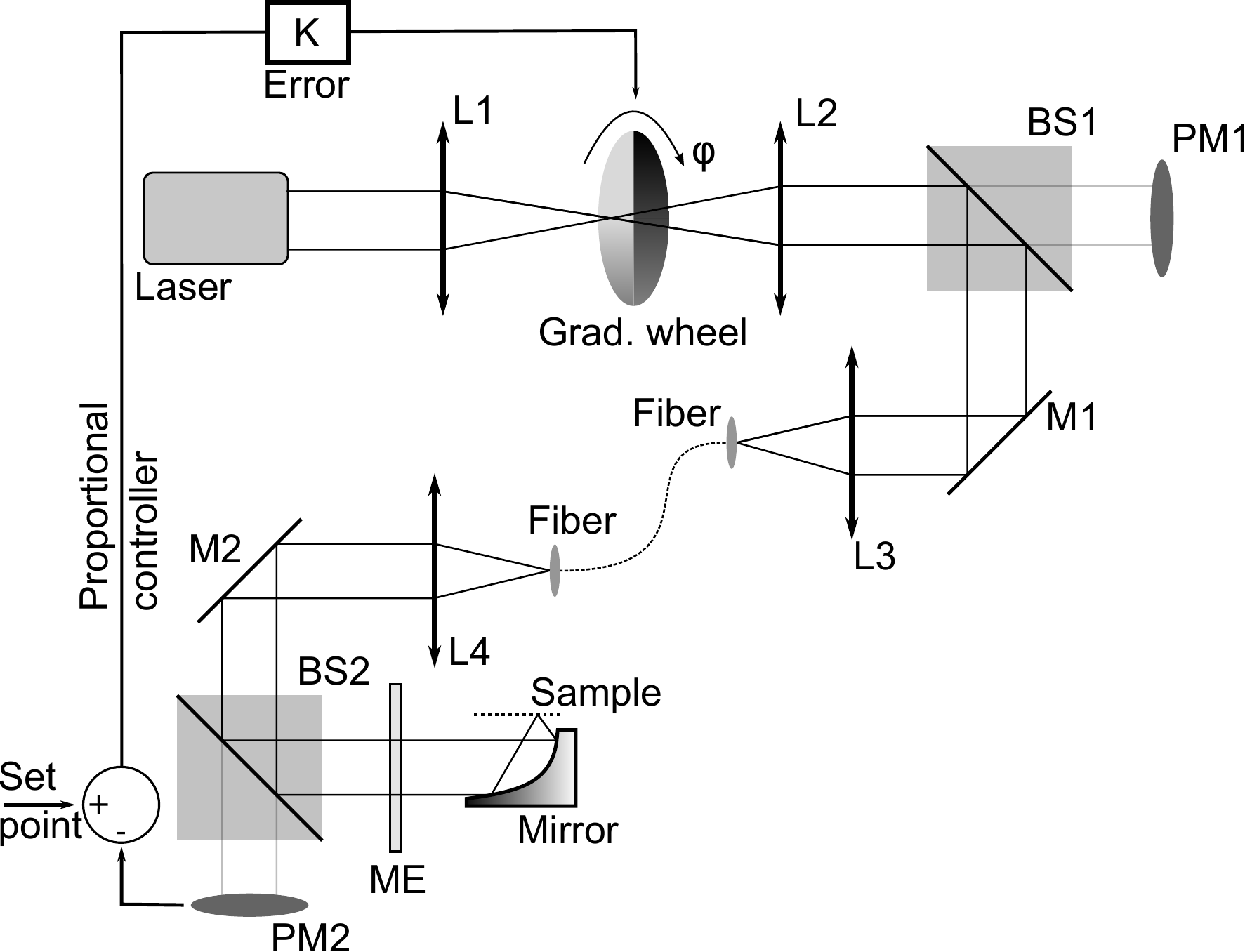}}
\caption{\textbf{Scheme of the optical components used in our EEGS experiment.} The laser output is focused on a gradient wheel density filter. The telescopic lenses L1 and L2 produce an output with unity magnification. The beam splitter BS1 is used to collect 10\% of the incident power in the power meter PM1. Finally, the laser beam is focused on the optical fiber. The other side of the output fiber is coupled to L4. M2 and BS2 are used to send the light beam to the parabolic mirror, focusing the beam on the sample plane (dashed line). The boundary between the atmosphere and vacuum (inside the microscope) is represented by the microscope edge ME. L1, L2: achromatic lenses; L3, L4: achromatic aspheric lenses; BS1, BS2: 90:10 beamsplitter; M1, M2: silver mirror; PM1, PM2: power meters.}
\label{fig:sup_optical_setup}
\end{figure}

% --- optical setup ---------------------------------------
\subsection{Details on the optical setup used to perform EEGS experiments}

The elements of the EEGS experimental setup are shown in Fig.~\ref{fig:sup_optical_setup}. The optical setup consists of two main parts: the coupling of the dye-laser light beam into the optical fiber (Fig.~\ref{fig:sup_optical_setup}a) and the other side of the optical fiber attached to the microscope light injection (or collection) system (Fig.~\ref{fig:sup_optical_setup}b). The achromatic lenses L1 and L2 in Fig.~\ref{fig:sup_optical_setup}a, which have the same focal point and, thus, form a telescopic imaging system with unity magnification, are used to produce an optical cross-over in which a continuous neutral density gradient wheel is placed. The latter is linked to a servo motor by a mechanical belt and two homemade 3D-printed rotating shafts. Additionally, an Arduino UNO is used to control the servo motor through the microcontroller standard library. The beamsplitter cube BS1 reflects 90\% of the incident light. The remaining 10\% is sent to the power meter PM1 (Thorlabs S120C sensor with Thorlabs PM100USB interface). The PM1 element is mostly used to determine the transmission percentage across the optical fiber, which is helpful during alignment and allow us to estimate the reminiscent chromatic aberration of the optical system. The silver mirror M1 and the achromatic aspheric lens L3 are used to focus the laser beam onto the fiber. In M1, there are two angular degrees of freedom controlled by piezoelectric actuators (Thorlabs PIAK10), while L3 is placed on a lens mount with 5 degrees of freedom (3 spatial and 2 angular). The fiber ends at a standard fiber SMA connector and is mounted on a 3-axis Thorlabs NanoMax\textsuperscript{TM} flexure stage using manual differential micrometers. Alignment is mainly carried out by playing with M1 and the fiber 3-axis stage. For a single-mode fiber, typical transmission values are $\sim$ 4-10\% across the Pyrromethene dye range (566--611\,nm wavelength), although these values are properly corrected by the gradient-wheel-based control system. To estimate the focusing power of the parabolic mirror, the field of view of the single-mode optical fiber is directly measured by rastering a $<100$\,nm diamond crystal containing multiple nitrogen-vacancy (NV) centers, which are known to efficiently emit light in the wavelength range of the dye laser used, thus reducing the effect of chromatic aberration. As shown in Fig.~\ref{fig:sup_optical_setup}c, when moving an NG center by approximately 0.8\,$\mu$m away from the optical axis of the electron microscope, the integrated collected signal is reduced by a factor of two.

% =========================================================
% --- EEGS theory -----------------------------------------
% =========================================================
\section{Analytical theory for the EEGS probability}

The interaction of a swift electron with a spherical object has previously been studied using analytical theory methods to calculate the EELS and cathodoluminescence (CL) emission probabilities \cite{paper149}, which should actually coincide when absorption by the probed materials is negligible (e.g., for the spheres considered in the present work). We derive below an analytical expression for the EEGS probability due to the interaction with an illuminated sphere by relating it to the CL far-field amplitude via the reciprocity theorem.

Considering a specimen subject to external monochromatic illumination of frequency $\omega$, the effect of the optical field on an electron moving with constant velocity $v$ along the $z$ direction is encapsulated in the coupling parameter \cite{paper371}
\begin{align}
\beta_{\rm EEGS}(\omega)=\frac{e}{\hbar\omega}\int_{-\infty}^\infty dz \, E_z(x_e,y_e,z) \, \ee^{-\ii\omega z/v},
\label{beta}
\end{align}
where $(x_e,y_e)$ defines the electron beam (e-beam) position in the transverse plane, while the amplitude $\Eb(\rb)$ is taken such that the time-dependent electric field reads $2{\rm Re}\big\{\Eb(\rb)\ee^{-\ii\omega t}\big\}$. We explicitly indicate the frequency dependence of $\beta_{\rm EEGS}(\omega)$ because this is a key ingredient in EEGS. After interaction, the incident zero-loss peak ($\ell=0$) splits into a series of peaks separated by multiples of the photon energy $\ell\hbar\omega$, with associated probabilities $P_\ell(\omega)=J_\ell^2(2|\beta_{\rm EEGS}(\omega)|)$ satisfying $\sum_{\ell=-\infty}^\infty P_\ell(\omega)=1$. In the present study, the coupling is relatively weak ($|\beta_{\rm EEGS}(\omega)|\ll1$), and therefore, only the first gain ($\ell=1$) and loss ($\ell=-1$) peaks are experimentally studied in the transmitted electron spectra, both of them having the same probability $J_1^2\big[2|\beta_{\rm EEGS}(\omega)|\big]\approx|\beta_{\rm EEGS}(\omega)|^2$.

% Figure S1 -----------------------------------------------
\begin{figure*}
\centering{\includegraphics[width=0.45\textwidth]{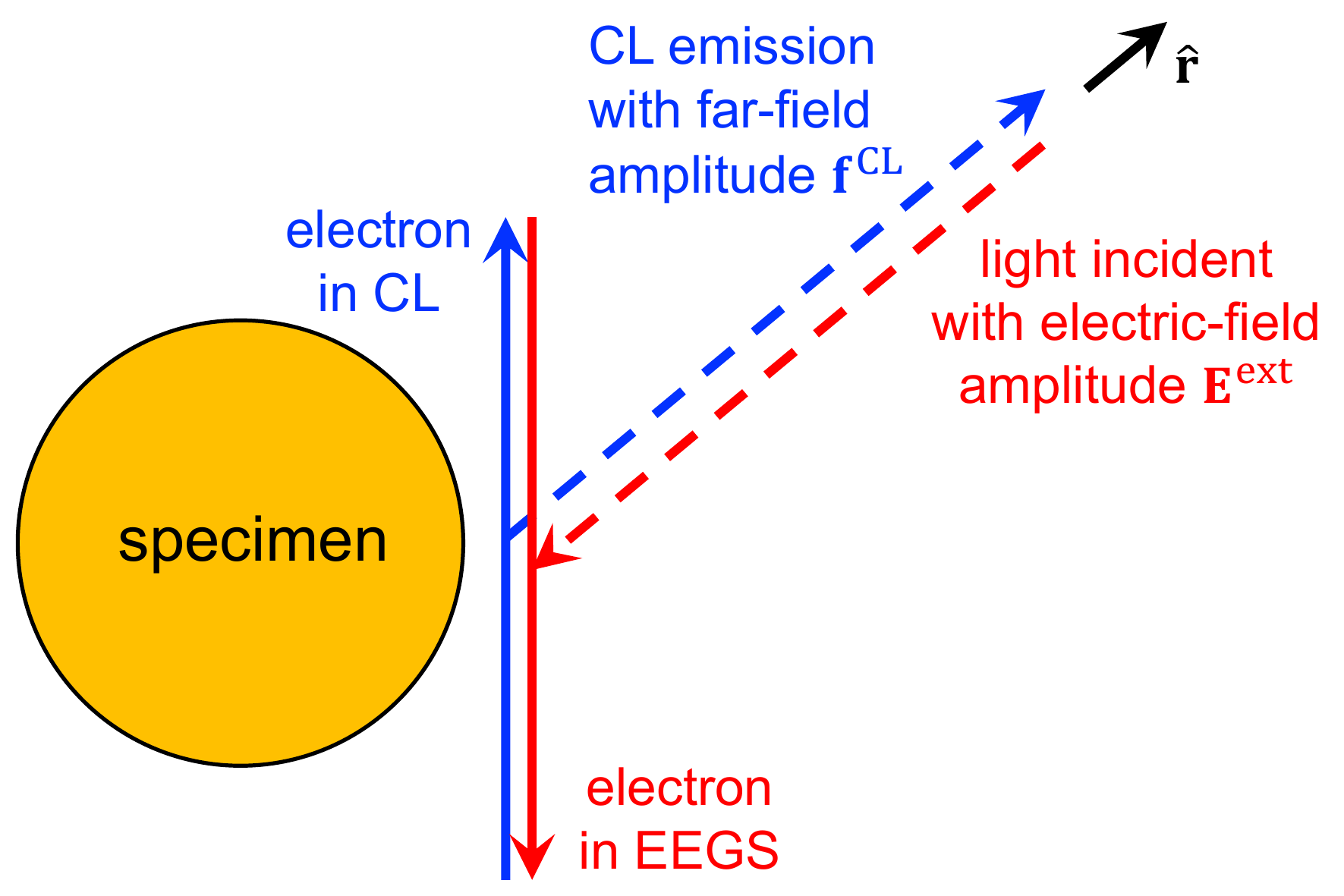}}
\caption{{\bf Sketch of the geometry involved in the application of the reciprocity theorem.} We relate the EEGS coupling coefficient $\beta_{\rm EEGS}(\omega)$ (red) to the CL far-field amplitude (blue) by means of Eq.~(\ref{betaf}). The propagation directions of electron and light are both reversed in CL relative to EEGS.}
\label{FigS1}
\end{figure*}

Rather than calculating $\beta_{\rm EEGS}(\omega)$ by first obtaining the total (incident+scattered) field $\Eb$ and then integrating Eq.~(\ref{beta}), we exploit the reciprocity theorem to relate $\beta_{\rm EEGS}(\omega)$ to the CL far-field amplitude ${\bf f}^{\rm CL}(\omega)$ as \cite{paper371}
\begin{align}
\beta_{\rm EEGS}(\omega)=\frac{\ii c^2}{\hbar\omega^2}{\bf f}^{\rm CL}_{\rr}(\omega)\cdot\Eb^{\rm ext},
\label{betaf}
\end{align}
where $\Eb^{\rm ext}$ is the EEGS incident electric field amplitude in the absence of the specimen. In this expression, ${\bf f}^{\rm CL}(\omega)$ needs to be computed after reversing the electron velocity vector relative to EEGS, and with the CL emission direction pointing to the EEGS light source, as indicated in Fig.~\ref{FigS1}. In particular, the CL emission direction is $\rr$, but the propagation direction of the EEGS incident light is $-\rr$ (see Fig.~\ref{FigS1}).

% --- betaEEGS for plane wave -----------------------------
\subsection{Spherical specimen under plane wave illumination}
\label{sphericalspecimen}

We now take a previously derived analytical expression for the CL far-field amplitude of a sphere in vacuum \cite{paper024}:
\begin{align}
{\bf f}^{\rm CL}_{\rr}(\omega)=\frac{c}{\omega}\sum_{l=1}^\infty\sum_{m=-l}^l \bigg[t_l^M \psi_{lm}^{M,{\rm ext}}\;\vec{\zeta}_{lm}(\rr)\,+ t_l^E \psi_{lm}^{E,{\rm ext}}\;\rr\times\vec{\zeta}_{lm}(\rr)\,\bigg],
\label{fCL}
\end{align}
where $l$ and $m$ run over multipolar components, $t_l^\nu$ are the $m$-independent Mie scattering coefficients for electric ($\nu=$E) and magnetic ($\nu=$M) polarization, and $\psi_{lm}^{\nu,{\rm ext}}$ are expansion coefficients that express the external evanescent field carried by the electron as a sum over multipoles around the sphere center. In Eq.~(\ref{fCL}), we use the vector spherical harmonics
\begin{align}
\vec{\zeta}_{lm}(\rr)=\frac{1}{2}\big[C_{lm}^-Y_{lm-1}(\rr)+C_{lm}^+Y_{lm+1}(\rr)\big]\,\xx\;
+\frac{\ii}{2}\big[C_{lm}^-Y_{lm-1}(\rr)-C_{lm}^+Y_{lm+1}(\rr)\big]\,\yy\;
+mY_{lm}(\rr)\,\zz,
\nonumber
\end{align}
where $C_{lm}^\pm=\sqrt{(l\pm m+1)(l\mp m)}$. Also, we consider the Mie scattering coefficients for a self-standing homogenous sphere of radius $a$ and permittivity $\epsilon_d$ centered at the origin, which are given by
\begin{align}
t_l^M &= \frac{- j_l(\rho_0) \rho_1 j_l^\prime(\rho_1)
+ \rho_0 j_l^\prime(\rho_0) j_l(\rho_1)}{h_l^{(+)}(\rho_0) \rho_1 j_l^\prime(\rho_1)
-\rho_0  [h_l^{(+)}(\rho_0)]^\prime j_l(\rho_1)}, \nonumber\\
t_l^E &= \frac{- j_l(\rho_0) \tilde{j}_l(\rho_1)^\prime
+ \epsilon_d \tilde{j}_l(\rho_0)^\prime j_l(\rho_1)}{h_l^{(+)}(\rho_0) \tilde{j}_l(\rho_1)^\prime
- \epsilon_d \tilde{h}_l^{(+)}(\rho_0)^\prime j_l(\rho_1)}, \nonumber
\end{align}
where $\rho_0=\omega a/c$, $\rho_1=\sqrt{\epsilon_d}\omega a/c$, $j_l$ and $h_l^{(+)}$ are spherical Bessel and Hankel functions \cite{M1966}, we have defined the functions $\tilde{j}_l(\rho)=\rho\, j_l(\rho)$ and $\tilde{h}_l^{(+)}(\rho)=\rho\, h_l^{(+)}(\rho)$, and the prime denotes differentiation with respect to the argument. Finally, the incident field supplied by the electron is represented through the coefficients \cite{paper027}
\begin{align}
\left[ \begin{array}{c}
\psi_{lm}^{M,{\rm ext}} \\ \\
\psi_{lm}^{E,{\rm ext}}
\end{array} \right]
=\frac{-2\pi\,e\,\ii^{1-l}\omega}{l(l+1)\,c^2} \, K_m\bigg[\frac{\omega R_e}{v\gamma}\bigg]\; \ee^{-\ii m\varphi_e} \,\times
\left[ \begin{array}{c}
2 m A_{lm} v/c     \\ \\
B_{lm}/\gamma
\end{array} \right],
\nonumber
\end{align}
where $(R_e,\varphi_e)$ are the polar coordinates of the electron transverse vector $(x_e,y_e)$, $K_m$ is the modified Bessel function of order $m$, $\gamma=1/\sqrt{1-(v/c)^2}$ is the Lorentz factor, and the coefficients $A_{lm}$ and $B_{lm}$ only depend on the normalized electron velocity $v/c$. More precisely \cite{WA1979},
\begin{align}
A_{lm} &= \ii^{l+m} (2 m-1)!! \,
\sqrt{\dfrac{(2 l+1)}{\pi}\dfrac{(l-m)!}{(l+m)!}}\,
\dfrac{(c/v)^{m+1}}{\gamma^m}\, C_{l-m}^{m+1/2}(c/v),
\;\;\;\;\; m\geq 0, \nonumber \\
B_{lm}  &=A_{l,m+1} \sqrt{(l+m+1)(l-m)} \,-\, A_{l,m-1} \sqrt{(l-m+1)(l+m)},
\nonumber
\end{align}
and $C^\mu_\nu$ are Gegenbauer polynomials \cite{AS1972}. We use the property $A_{l,-m}=(-1)^m A_{lm}$ to calculate the $A_{lm}$ coefficients for negative $m$.

These results allow us to calculate the EEGS probability $|\beta_{\rm EEGS}(\omega)|^2$ for an electron moving downwards and passing close to a homogeneous sphere under plane-wave light irradiation from Eq.~(\ref{betaf}) by using the CL amplitude ${\bf f}_{\rr}^{\rm CL}(\omega)$ given by Eq.~(\ref{fCL}) for an electron moving along the positive $z$ axis (see Fig.~\ref{FigS1}).

% --- parabolic mirror ------------------------------------
\subsection{Illumination by means of a parabolic mirror}
\label{parabolicmirror}

Illumination of the specimen in the microscope is introduced through a parabolic aluminum mirror, producing a focal spot that can be regarded as a superposition of plane waves resulting from reflection of the incident laser on different regions of the metal surface. We model the system by describing the reflection from each mirror surface position $\sb$ through the Fresnel coefficients of a tangent planar mirror at that position.

% Figure S2 -----------------------------------------------
\begin{figure*}
\centering{\includegraphics[width=0.65\textwidth]{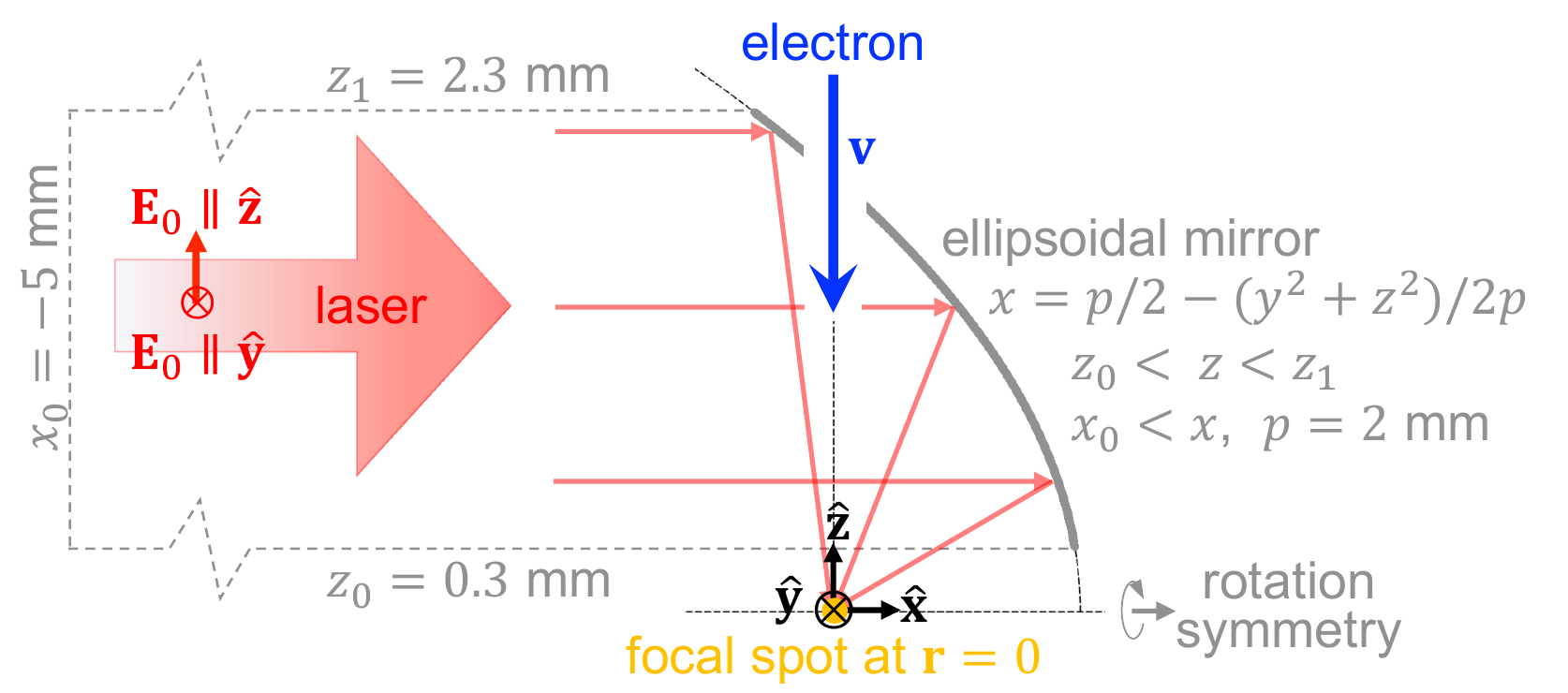}}
\caption{{\bf Illumination configuration used in the present work.} We assume light incident from the left and propagating towards the positive $x$ direction, the e-beam is injected through a hole and moves downwards along the negative $z$ direction, and the specimen is placed near the focal spot at $\rb=0$. The mirror surface is defined by the equation $x=p/2-(y^2+z^2)/2p$. In the present experiment, we have $p=2\,$mm, and the mirror surface is limited by the conditions $x>x_0=-5\,$mm and $z_0=0.3\,$mm\,$<z<z_1=2.3\,$mm. The hole is ignored in our calculations. In this work, we consider a laser field amplitude $\Eb_0$ oriented along either the $y$ or $z$ directions.}
\label{FigS2}
\end{figure*}

In what follows, we consider the e-beam to move along the negative $z$ direction and the parabolic mirror to be illuminated by a laser plane wave propagating along the positive $x$ direction, as illustrated in Fig.~\ref{FigS2}. To obtain the field reaching the specimen region after reflection by the mirror, we assimilate the response of each surface element to effective $\sb$-dependent electric and magnetic dipole densities $\pb_\sb$ and $\mb_\sb$, the magnitudes of which are obtained in the limit of a planar mirror, as we mentioned above.

We now analyze the equivalent dipole sources produced by reflection on a planar mirror. More precisely, for an incident light plane wave $\Eb^i\ee^{\ii\kb^i\cdot\rb}$ of wave vector $\kb^i$ and electric field amplitude $\Eb^i$, the field reflected by a planar mirror that contains the origin $\rb=0$ and has an outer normal unit vector $\nn$ is given by $\Eb^r\ee^{\ii\kb^r\cdot\rb}$, where $\kb^r=\kb^i-2(\kb^i\cdot\nn)\nn$ is the reflected wave vector, and $\Eb^r=\rs(\Eb^i\cdot\eh_s)\,\eh_s+\rp(\Eb^i\cdot\eh_p^i)\,\eh_p^r$ is the reflected field amplitude written in terms of the Fresnel coefficients $\rs$ and $\rp$ for $s$ and $p$ polarization. Here, $\eh_s$ and $\eh_p^\nu$ are the corresponding unit polarization vectors. We note that the $p$-polarization vector $\eh_p^\nu$ is different for incidence ($\nu=i$) and reflection ($\nu=r$) directions (see below). Translational invariance of the planar mirror allows us to incorporate the $\sb$ dependence of the effective dipoles through $\pb_\sb=\pb\,\ee^{\ii\kparb\cdot\sb}$ and $\mb_\sb=\mb\,\ee^{\ii\kparb\cdot\sb}$, where $\kparb=\kb^i-(\kb^i\cdot\nn)\nn$ is the in-plane component of the incident wave vector. In terms of these effective sources, and assuming a permittivity $\epsilon_0$ ($=1$ in the present work) in the medium outside the mirror, the reflected field becomes
\begin{align}
\int d^2\sb\;\Big[k^2\,\pb_\sb+\frac{1}{\epsilon_0}(\pb_\sb\cdot\nabla_\rb)\nabla_\rb-\ii k\,\mb_\sb\times\nabla_\rb\Big] \,\dfrac{\ee^{\ii k\sqrt{\epsilon_0}|\rb-\sb|}}{|\rb-\sb|}=\frac{2\pi\ii k^2}{k_\perp}\,\ee^{\ii\kb^r\cdot\rb}\,\Big[\pb-(\pb\cdot\hat{\kb}^r)\hat{\kb}^r+ \sqrt{\epsilon_0}\;\mb\times\hat{\kb}^r\Big],
\label{Eref1}
\end{align}
where $k=\omega/c$, $k_\perp=-\kb^i\cdot\nn$, and the $\sb$ integral is extended over the plane normal to $\nn$. Comparing this result with the reflected field $\Eb^r\ee^{\ii\kb^r\cdot\rb}$ and noticing that $\Eb^r\cdot\kb^r=0$, we find that we can take $\mb=0$ and $\pb=(-\ii k_\perp/2\pi k^2)\,\Eb^r$. Finally, inserting the expression given above for $\Eb^r$ in terms of the Fresnel coefficients, we have $\pb=(-\ii k_\perp/2\pi k^2)\,\big[\rs(\Eb^i\cdot\eh_s)\,\eh_s+\rp(\Eb^i\cdot\eh_p^i)\,\eh_p^r\big]$.

For a parabolic mirror defined by $x=p/2-(y^2+z^2)/2p$ (focal distance $p/2$), as shown in Fig.~\ref{FigS2}, we define the coordinates $\rb_\perp=(y,z)$ in the transverse plane relative to the rotation axis $x$ and write the elements involved in the calculation of the reflected field at each position $\sb$ as
\begin{subequations}
\label{various}
\begin{align}
&\nn=-(\rb_\perp+p\xx)/\sqrt{r_\perp^2+p^2}, \\
&\hat{\kb}^i=\xx, \\
&\hat{\kb}^r=[(r_\perp^2-p^2)\xx-2p\rb_\perp)]/(r_\perp^2+p^2), \\
&\eh_s=(-z\yy+y\zz)/r_\perp, \\
&\eh_p^i=\hat{\rb}_\perp, \\
&\eh_p^r=[(r_\perp^2-p^2)\rb_\perp+2pr_\perp\xx)]/(r_\perp^2+p^2),
\end{align}
\end{subequations}
and $\kb^\nu=k\sqrt{\epsilon_0}\,\hat{\kb}^\nu$. In addition, the normal wave vector becomes $k_\perp=k\sqrt{\epsilon_0}\,p/\sqrt{r_\perp^2+p^2}$. Using these elements, we can calculate the electric field $\Eb(\rb)$ produced by the mirror at a position $\rb$ near the focal point ($\rb=0$) from an expression similar to the left-hand side of Eq.~(\ref{Eref1}), but now with $\sb$ integrated over the surface of the curved mirror. More precisely,
\begin{subequations}
\label{Eref2}
\begin{align}
\Eb(\rb_f)&=\int_{\rm mirror}\!\!\!\!\!\!\!\! d^2\sb\;\Big[k^2\,\pb_\sb+\frac{1}{\epsilon_0}(\pb_\sb\cdot\nabla_{\rb_f})\nabla_{\rb_f}\Big] \,\dfrac{\ee^{\ii k\sqrt{\epsilon_0}|\rb_f-\sb|}}{|\rb_f-\sb|} \label{Eref2a}\\
&\approx \ee^{\ii k p\sqrt{\epsilon_0}}\int_{\rm mirror}\!\!\!\!\!\!\!\! d^2\sb\;\Big[k^2\,\pb_\sb-\frac{1}{\epsilon_0}(\pb_\sb\cdot\kb^r)\kb^r\Big] \,\dfrac{\ee^{\ii \kb^r\cdot\rb_f}}{s} \label{Eref2b}\\
&=-\ee^{\ii k p\sqrt{\epsilon_0}}\,\frac{\ii}{2\pi}\int_{\rm mirror}\!\!\!\!\!\!\!\! d^2\sb\; \frac{k_\perp}{s}\,\ee^{\ii \kb^r\cdot\rb_f}\,\big[\rs(\Eb_0\cdot\eh_s)\,\eh_s+\rp(\Eb_0\cdot\eh_p^i)\,\eh_p^r\big] \label{Eref2c}\\
&=-\ee^{\ii k p\sqrt{\epsilon_0}}\,\frac{\ii k\sqrt{\epsilon_0}}{2\pi}\int_{\rm mirror}\!\!\!\!\! d\rb_\perp\; \frac{1}{s}\,\ee^{\ii \kb^r\cdot\rb_f}\,\big[\rs(\Eb_0\cdot\eh_s)\,\eh_s+\rp(\Eb_0\cdot\eh_p^i)\,\eh_p^r\big], \label{Eref2d}
\end{align}
\end{subequations}
where (\ref{Eref2b}) is obtained from (\ref{Eref2a}) by adopting the $kp\gg1$ limit (i.e., considering that the focal distance $p/2$ is large compared with the light wavelength) and using the relation $\hat{\kb}^r=-\hat{\sb}$ to connect the reflected wave vector to the direction of the surface position $\sb$ (as seen from the focus) in the integral. In doing so, we are assuming an incident plane wave $\Eb_0\,\ee^{\ii kx}$ illuminating the mirror, such that the sum of light propagation distances from the $x=0$ plane to the mirror surface and from this to the focal point amounts to a path length given by $p$ and emerging as an irrelevant global phase factor in Eq.~(\ref{Eref2b}). Then, Eq.~(\ref{Eref2c}) is obtained by noticing that $\pb_\sb\cdot\kb^r=0$ because $\pb_\sb\propto\Eb^r$ and the reflected field must be transverse. In addition, we insert the explicit expression for $\pb$ derived above for the planar mirror in terms of the Fresnel coefficients, with $\Eb^i$ substituted by $\Eb_0$. We note that the $\sb$-dependent phase in $\Eb^i$ is already included in the global phase factor $\ee^{\ii k p\sqrt{\epsilon_0}}$. Finally, Eq.~(\ref{Eref2d}) results from a change of integration variables $\sb\to\rb_\perp$ after noticing that the corresponding Jacobian times $k_\perp$ reduces to $k\sqrt{\epsilon_0}$. The electric field in the focal region is then given by Eq.~(\ref{Eref2d}) using the quantities defined in  Eqs.~(\ref{various}) in terms of $\rb_\perp=(y,z)$ as well as $s=(p^2+r_\perp^2)/2p$. Also, we use the Fresnel coefficients
\begin{subequations}
\label{Fresnel}
\begin{align}
&\rs=(k_\perp-k'_\perp)/(k_\perp+k'_\perp), \\
&\rp=(\epsilon_mk_\perp-\epsilon_0k'_\perp)/(\epsilon_mk_\perp+\epsilon_0k'_\perp),
\end{align}
\end{subequations}
where $k'_\perp=\sqrt{k^2(\epsilon_m-\epsilon_0)+k_\perp^2}$ and $\epsilon_m$ is the metal permittivity.

% Figure S3 -----------------------------------------------
\begin{figure*}
\centering{\includegraphics[width=0.6\textwidth]{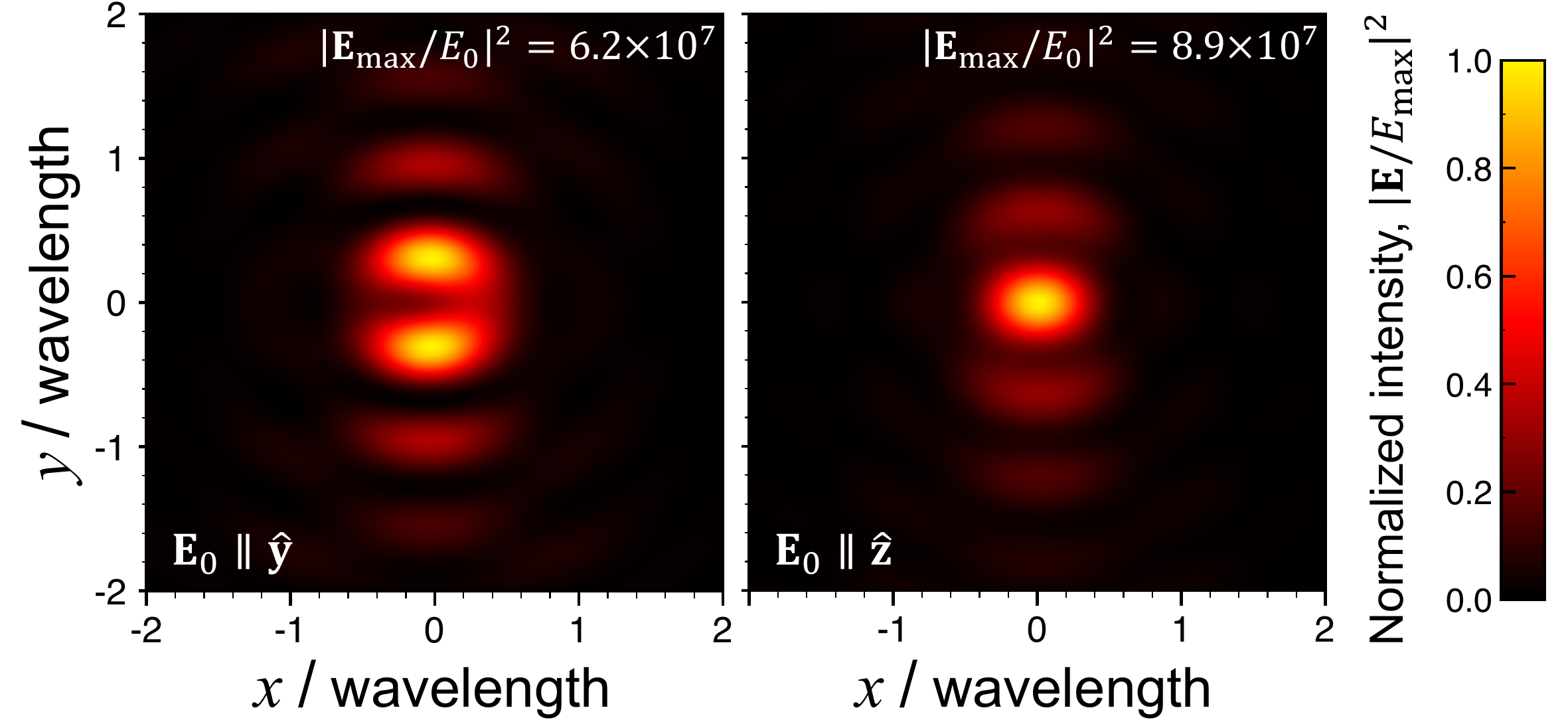}}
\caption{{\bf Focal spot produced in the $z=0$ plane by the mirror in Fig.~\ref{FigS2}.} We consider incident light polarization along either the $y$ (left) or the $z$ (right) directions. The electric field intensity is normalized to the maximum $|E_{\rm max}|^2$, with values of the latter relative to the incident field intensity $|E_0|^2$ also indicated. The calculation is made for an aluminum mirror at a photon energy of 2\,eV (i.e., a wavelength $\lambda=620\,$nm). Similar results are obtained when assimilating the metal to a perfect conductor, so these results describe well the focal spot at frequencies up to the visible range. The mirror has a projected area on the $yz$ plane of 18.74\,mm$^2=4.87\times10^7\,\lambda^2$. The effective projected area including absorption in aluminum is 16.51\,mm$^2$ and 17.35\,mm$^2$ for $\Eb_0\parallel\yy$ and $\Eb_0\parallel\zz$, respectively.}
\label{FigS3}
\end{figure*}

In our experimental setup, we have an aluminum mirror ($\epsilon_m(\omega)$ taken from optical data \cite{P1985}) in vacuum ($\epsilon_0=1$) with a focal distance of 1\,mm (i.e., $p=2\,$mm) and the metal surface limited by the conditions stated in Fig.~\ref{FigS2}. We adopt these parameters throughout the present paper. Then, the focal spot size is about 1$\,\mu$m, with the actual shape depending on the polarization of the incident light, as shown in Fig.~\ref{FigS3}.

% --- betaEEGS for a mirror -------------------------------
\subsection{EEGS under illumination by a parabolic mirror}

Like the focal spot, the electron--light coupling coefficient $\beta_{\rm EEGS}(\omega)$ can also be regarded as a superposition of contributions due to different reflected plane waves. Using Eq.~(\ref{betaf}) for the contribution of each of these waves, integrating over the mirror surface just like in Eq.~(\ref{Eref2d}), and ignoring any global phase, we find
\begin{subequations}
\label{betaEEGSfinal}
\begin{align}
\beta_{\rm EEGS}(\omega)=\frac{c}{\pi\hbar\omega}\int_{\rm mirror}\!\!\!\!\! d\rb_\perp\;&\frac{p}{p^2+r_\perp^2}\;{\bf f}^{\rm CL}_{\hat{\sb}}(\omega)\cdot\big[\rs(\Eb_0\cdot\eh_s)\,\eh_s+\rp(\Eb_0\cdot\eh_p^i)\,\eh_p^r\big] \label{betaEEGSfinala}\\
=\frac{c}{\pi\hbar\omega}\int_{z_0}^{z_1}dz\int_{-y_0(z)}^{y_0(z)}dy\;&\frac{p}{p^2+r_\perp^2}\;{\bf f}^{\rm CL}_{\hat{\sb}}(\omega)\cdot\big[\rs(\Eb_0\cdot\eh_s)\,\eh_s+\rp(\Eb_0\cdot\eh_p^i)\,\eh_p^r\big],
\label{betaEEGSfinalb}
\end{align}
\end{subequations}
involving $\hat{\sb}=-\hat{\kb}^r$ as well as different quantities defined in Eqs.~(\ref{various}) and (\ref{Fresnel}). We remark that Eq.~(\ref{betaEEGSfinala}) is general for any parabolic mirror geometry, while Eq.~(\ref{betaEEGSfinalb}) is specialized to our mirror dimensions, as described in Fig.~\ref{FigS2}, such that the limits of integration are $z_0=0.3\,$mm, $z_1=2.3\,$mm, and $y_0(z)=\sqrt{p^2-2px_0-z^2}$ with $x_0=-5\,$mm and $p=2\,$mm. Also, Eqs.~(\ref{betaEEGSfinal}) can be applied to any specimen by plugging the corresponding CL amplitude ${\bf f}^{\rm CL}_{\hat{\sb}}(\omega)$. In the present study, we calculate the latter for a sphere by means of Eq.~(\ref{fCL}).

% Figure S4 -----------------------------------------------
\begin{figure*}
\centering{\includegraphics[width=0.7\textwidth]{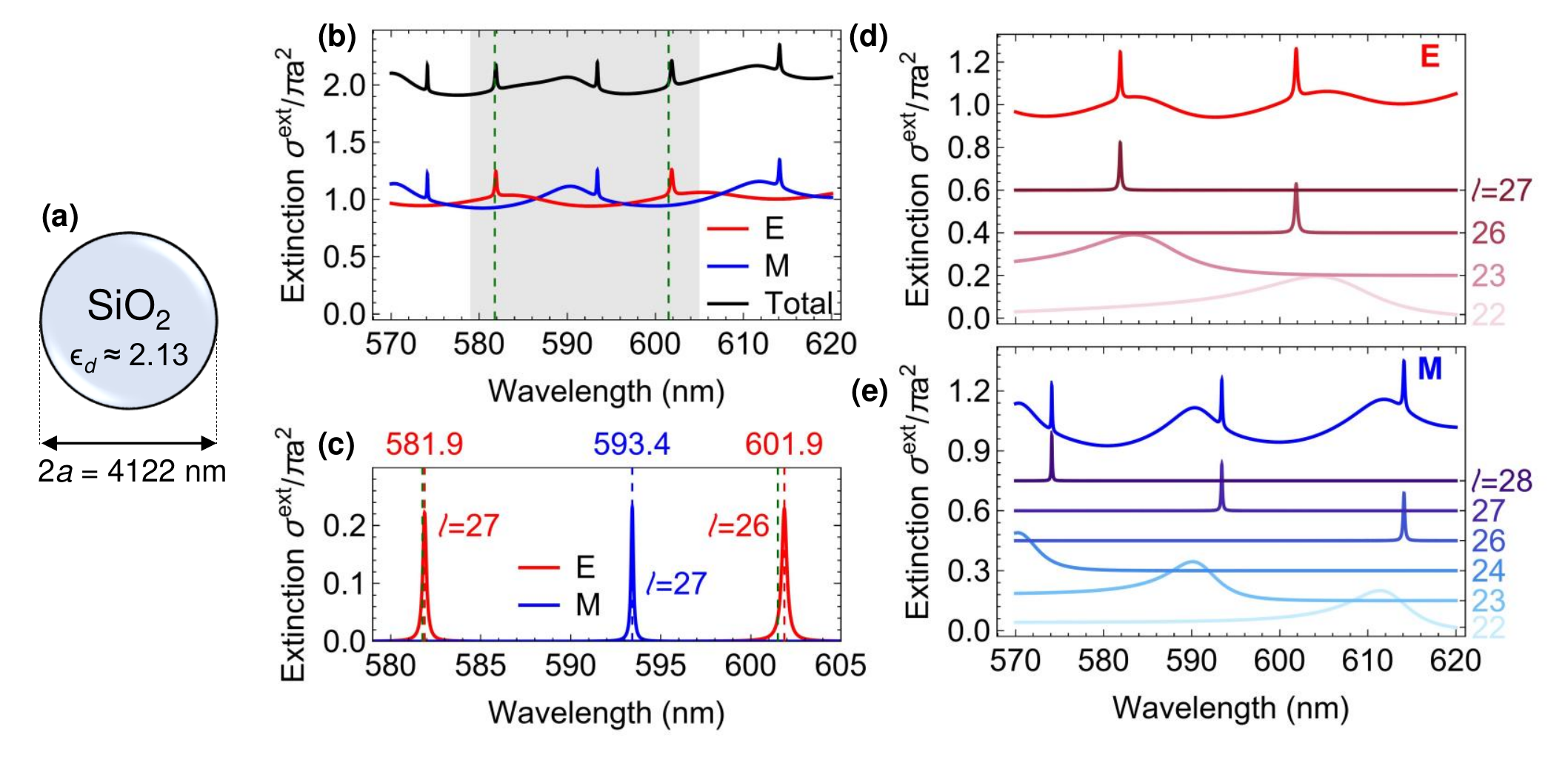}}
\caption{{\bf Extinction cross-section spectrum of a large SiO$_2$ sphere.} {\bf (a)} Sketch of the sphere with a diameter $2a=4122\,$nm and a permittivity $\epsilon_d\approx2.13$. {\bf (b)} Total optical extinction cross section $\sigma^{\rm ext}$ (black curve) normalized to the geometrical cross section $\pi a^2$, along with the corresponding partial contributions due to electric (red) and magnetic (blue) modes, as a function of light wavelength. The vertical dashed lines indicate the two resonance wavelengths observed in the experimental results of Fig.~2 in the main text. {\bf (c)} Partial extinction cross sections within the shaded region of panel (b), allowing us to identify two electric modes with $l=26$ and $27$ (red), as well as a magnetic mode with $l=27$ (blue). The vertical dashed green lines are the same as in panel (b). {\bf (d)} Total extinction cross section due to electric modes (top curve) and partial contributions from different multipolar orders $l$ (see labels on the right axis). A vertical offset between different curves is introduced for clarity. {\bf (e)} Same as panel (d), but for magnetic modes.}
\label{FigS4}
\end{figure*}

% Figure S5 -----------------------------------------------
\begin{figure*}
\centering{\includegraphics[width=0.7\textwidth]{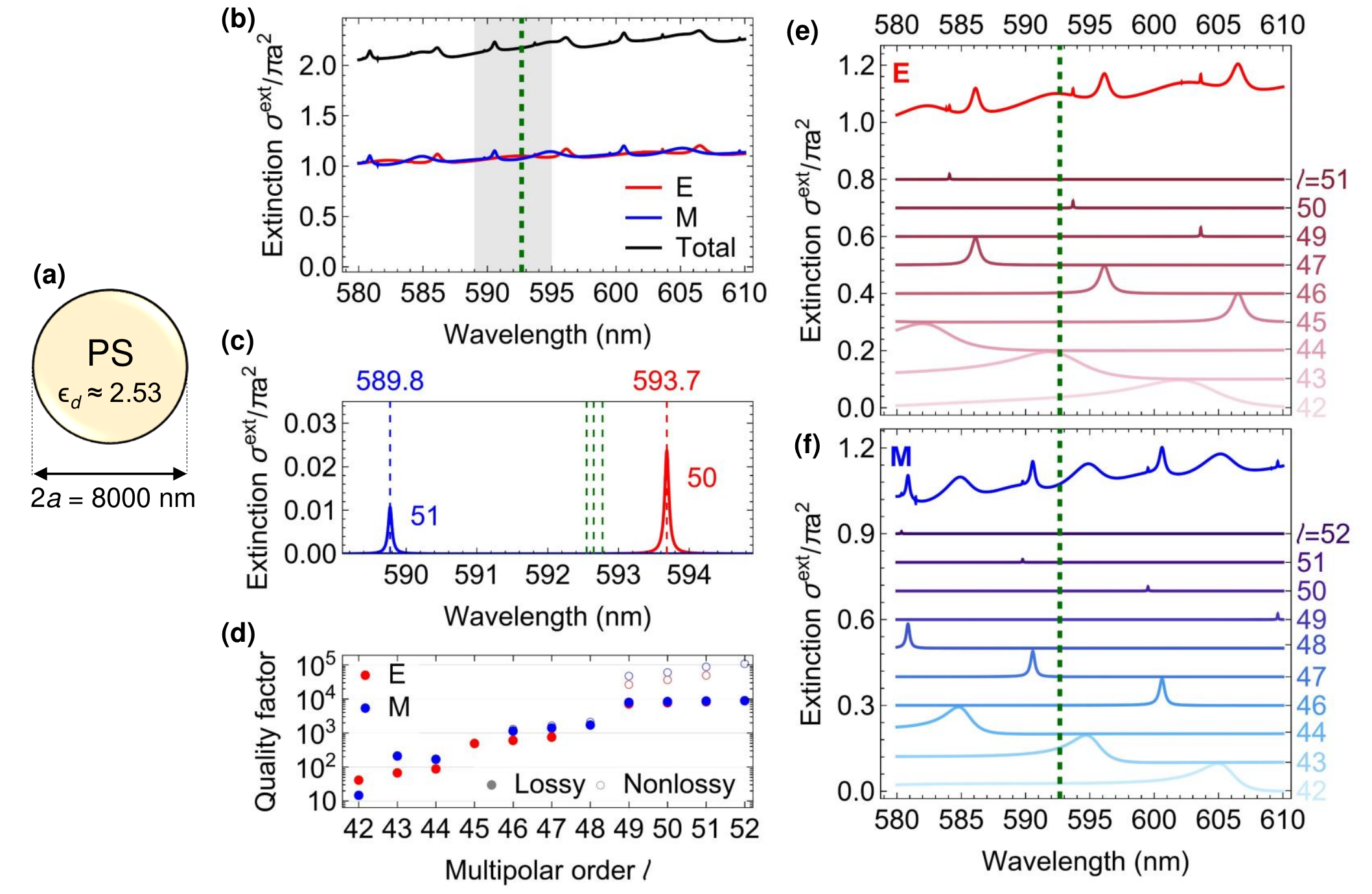}}
\caption{{\bf Extinction cross-section spectrum of a large polystyrene (PS) sphere.} {\bf (a)} Sketch of the sphere with a diameter $2a=8000\,$nm and a permittivity $\epsilon_d\approx2.53\times(1+\ii\eta)$, where losses are phenomenologically introduced through a small imaginary part $\eta=10^{-4}$ in the multiplying factor. {\bf (b)} Total optical extinction cross section $\sigma^{\rm ext}$ (black curve) normalized to the geometrical cross section $\pi a^2$, along with the corresponding partial contributions due to electric (red) and magnetic (blue) modes, as a function of light wavelength. The vertical dashed lines indicate the position of the resonances observed in the experimental measurements (see Fig.~3 in the main text). {\bf (c)} Partial cross sections contributed by magnetic modes with $l=46,47$ (blue) and electric modes with $l=45$-47 (red), giving rise to the resonances observed in panel (b). The vertical dashed green lines are the same as in panel (b). {\bf (d)} Quality factors of electric (red) and magnetic (blue) Mie resonances as a function of their multipolar order $l$. Solid and open circles are calculated with $\eta=10^{-4}$ (partially lossy PS) and $\eta=0$ (nonlossy PS), respectively. {\bf (e)} Total extinction cross section due to electric modes (top curve) and partial contributions from different multipolar orders $l$ (see labels on the right axis). A vertical offset between different curves is introduced for clarity. {\bf (f)} Same as panel (e), but for magnetic modes.}
\label{FigS5}
\end{figure*}
 
% =========================================================
% --- Results: Mie modes ----------------------------------
% =========================================================
\section{Mie modes in large dielectric spheres}
\label{Miemodes}

To characterize the large dielectric spheres studied in this work, we first calculate their optical extinction cross sections as a function of light wavelength $\lambda=2\pi c/\omega$ from Mie theory \cite{M1908} using the expression
\begin{align}
\sigma^{\rm ext}(\lambda)=\frac{\lambda}{2\pi\sqrt{\epsilon_0}}\sum_{l=1}^\infty(2l+1) \big[{\rm Im}\{t_l^M\}+{\rm Im}\{t_l^E\}\big],
\nonumber
\end{align}
where we set $\epsilon_0=1$ (particle in vacuum).

For the $4122\,$nm silica sphere, we fine-tune the diameter to match the two prominent resonances observed in the EEGS/EELS/CL experiments. The results presented in Fig.~\ref{FigS4} show that such resonances have electric polarization with multipolar orders $l=27$ and $26$ at wavelengths of $\sim582\,$nm and $\sim602\,$nm, respectively. There is an additional magnetic resonance at $\sim593\,$nm with $l=27$, which is missed in the experiments for symmetry reasons, as we argue below. The quality factors of these modes are $\sim1500$, substantially larger than those observed in experiment $\sim200$, in which losses at the support or due to surface contamination can affect the actual resonances.

A larger polysterene (PS) sphere is also studied, for which we predict modes with quality factors reaching $10^6$ when the particle is self-standing (Fig.~\ref{FigS5}d). This is in contrast to the values $\sim10^4$ observed in experiment, a disagreement that could be attributed to losses originating in the effect of the supporting film. In addition, we cannot find a good match between the calculated Mie resonances and the three peaks reported in Fig.~4 of the main text even when fine-tuning the sphere diameter. A possible reason for this discrepancy is that the particle (including the effect of the support) deviates from the spherical geometry assumed in Mie theory, such that the three observed peaks originate in the splitting of a single high-order multipolar mode relative of the idealized sphere. We therefore search for a possible Mie resonance in the sphere that has large optical strength in the measured spectral range. To compare with experiment, we introduce a factor $1+10^{-4}\,\ii$ in the material permittivity, so that the maximum quality factors of the Mie resonances are brought down to $\sim10^4$ (see Fig.~\ref{FigS5}d). The resulting multipolar decomposition of this slightly lossy sphere reveals a magnetic resonance of multipolar order $l=51$ and an electric one of order $l=50$ near the observed features (Fig.~\ref{FigS5}c), so these are possible candidates to explain the measurements if a small shift and splitting is produced by the lack of sphericity or the effect of the support. Such effects would be less relevant in the smaller silica sphere studied above because of the larger width of its resonances.

% Figure S6 -----------------------------------------------
\begin{figure*}
\centering{\includegraphics[width=0.5\textwidth]{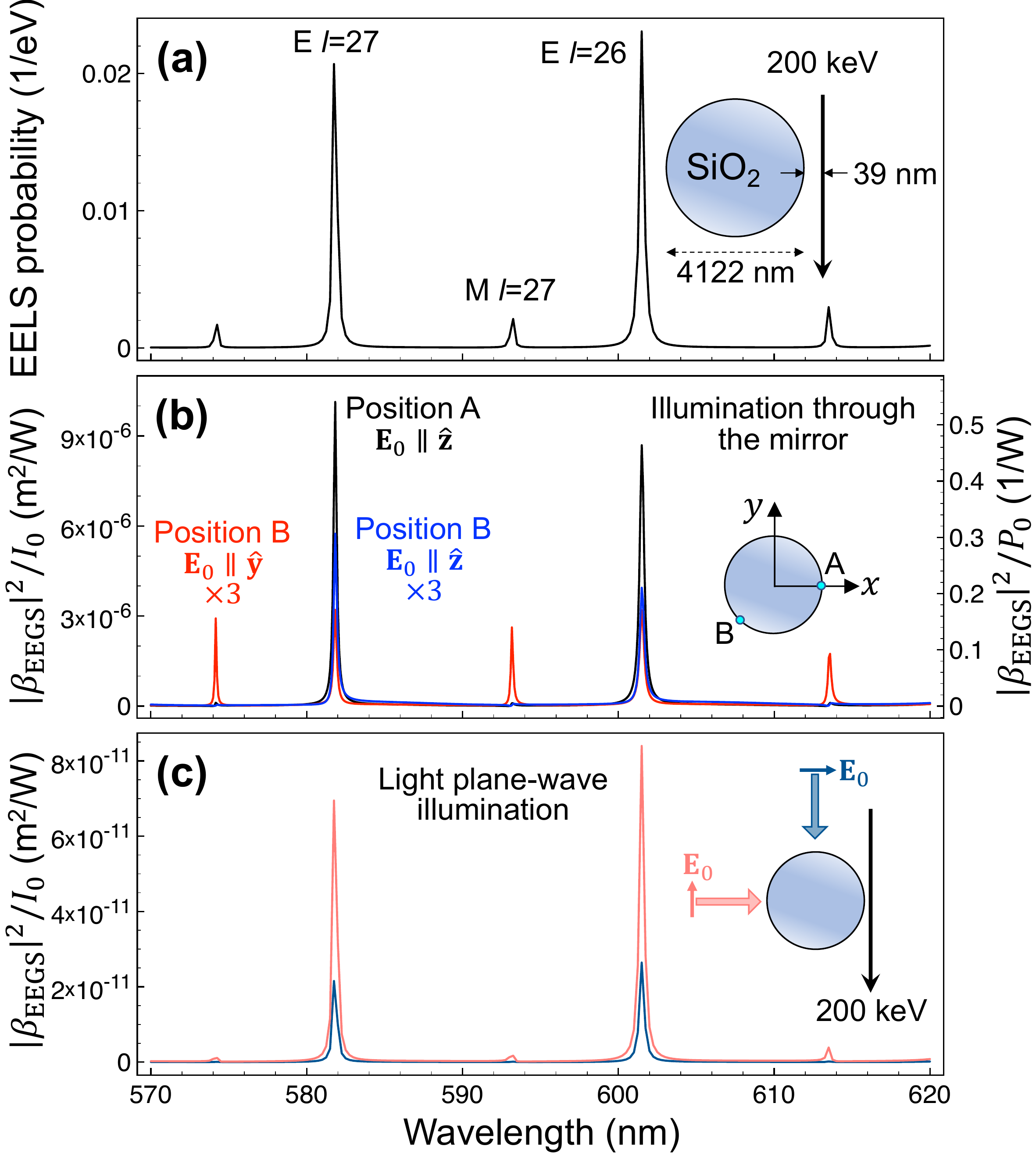}}
\caption{{\bf EELS and EEGS in the silica sphere of Fig.~\ref{FigS4}}. {\bf (a)} EELS probability calculated from Eq.~(\ref{GammaEELS}) as a function of light wavelength $2\pi c/\omega$. {\bf (b)} EEGS probability $|\beta_{\rm EGGS}(\omega)|^2$ obtained from Eq.~(\ref{betaEEGSfinalb}) and normalized either to the laser intensity $I_0$ that is incident on the mirror described in Fig.~\ref{FigS2} ($|\beta_{\rm EGGS}(\omega)|^2/I_0$, left vertical scale) or to the laser power $P_0$ ($|\beta_{\rm EGGS}(\omega)|^2/P_0$, right vertical scale, with $P_0=I_0A$, where $A=18.74\,$mm$^2$ is the area of the mirror projection on the $yz$ plane), with the laser focal spot aimed at positions that are 57\,nm inside the projected sphere contour; the EEGS probability is calculated for two different azimuthal locations of the focal spot and the e-beam (A and B in the inset, with the focal spot and the e-beam sharing the same azimuthal position in each case), it vanishes in A for incident light polarized along $y$ due to symmetry considerations, and it is multiplied by a factor of 3 in B. The orientation of the Cartesian axes is the same as in Fig.~\ref{FigS2}. {\bf (c)} EEGS probability computed from Eq.~(\ref{betaf}) and normalized to the incident light intensity for plane-wave illumination with two different directions of light incidence (see color-coordinated thick arrows in the inset). Electrons have an energy of 200\,keV and pass at a distance of 39\,nm from the sphere surface in all cases.}
\label{FigS6}
\end{figure*}

% Figure S7 -----------------------------------------------
\begin{figure*}
\centering{\includegraphics[width=0.5\textwidth]{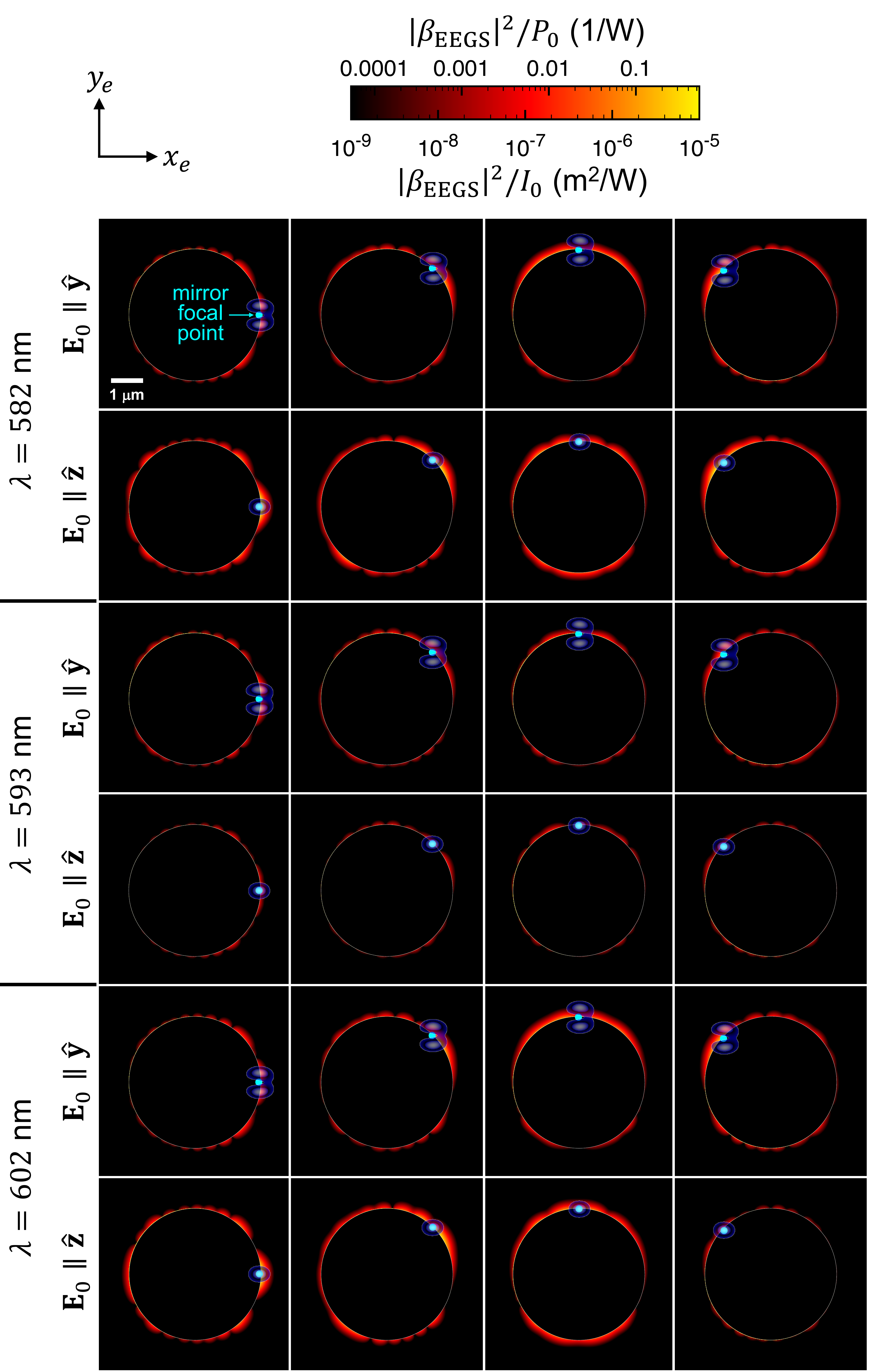}}
\caption{{\bf Spatial maps of the EEGS probability for the silica sphere of Fig.~\ref{FigS4} (4122\,nm diameter).} We plot $|\beta_{\rm EGGS}(\omega)|^2$ as a function of e-beam position $(x_e,y_e)$ (see upper-left frame, coordinated with Fig.~\ref{FigS2}) for illumination on resonance with the three Mie modes of Fig.~\ref{FigS4}c (see the corresponding wavelengths $\lambda=2\pi c/\omega$ on the left) and 200\,keV electron energy. We consider two different orientations of the incident field $\Eb_0$ referred to the mirror in Fig.~\ref{FigS2} ($\parallel\yy$ or $\parallel\zz$, see labels on the left). Results are presented for four different positions of the laser focal spot (blue dots) relative to the sphere (white contours). The EEGS probability $|\beta_{\rm EGGS}(\omega)|^2$ (color log scale) is normalized to the laser intensity $I_0=c|E_0|^2/2\pi$ incident on the mirror (i.e., $|\beta_{\rm EGGS}(\omega)|^2/I_0$, lower scale) and the laser power $P_0=I_0A$ incident on the mirror (i.e., $|\beta_{\rm EGGS}(\omega)|^2/P_0$, upper scale), where $A=18.74\,$mm$^2$ is the area of the mirror projection on the $yz$ plane. We only plot $|\beta_{\rm EGGS}(\omega)|^2$ for e-beam positions that do not intersect the particle. The focal-spot intensity at $z=0$ (taken from Fig.~\ref{FigS3}) is superimposed for each $\Eb_0$ orientation (color scale from maximum intensity (blue) down to a fraction $1/e$ of it (white contour)).}
\label{FigS7}
\end{figure*}

% Figure S8 -----------------------------------------------
\begin{figure*}
\centering{\includegraphics[width=0.5\textwidth]{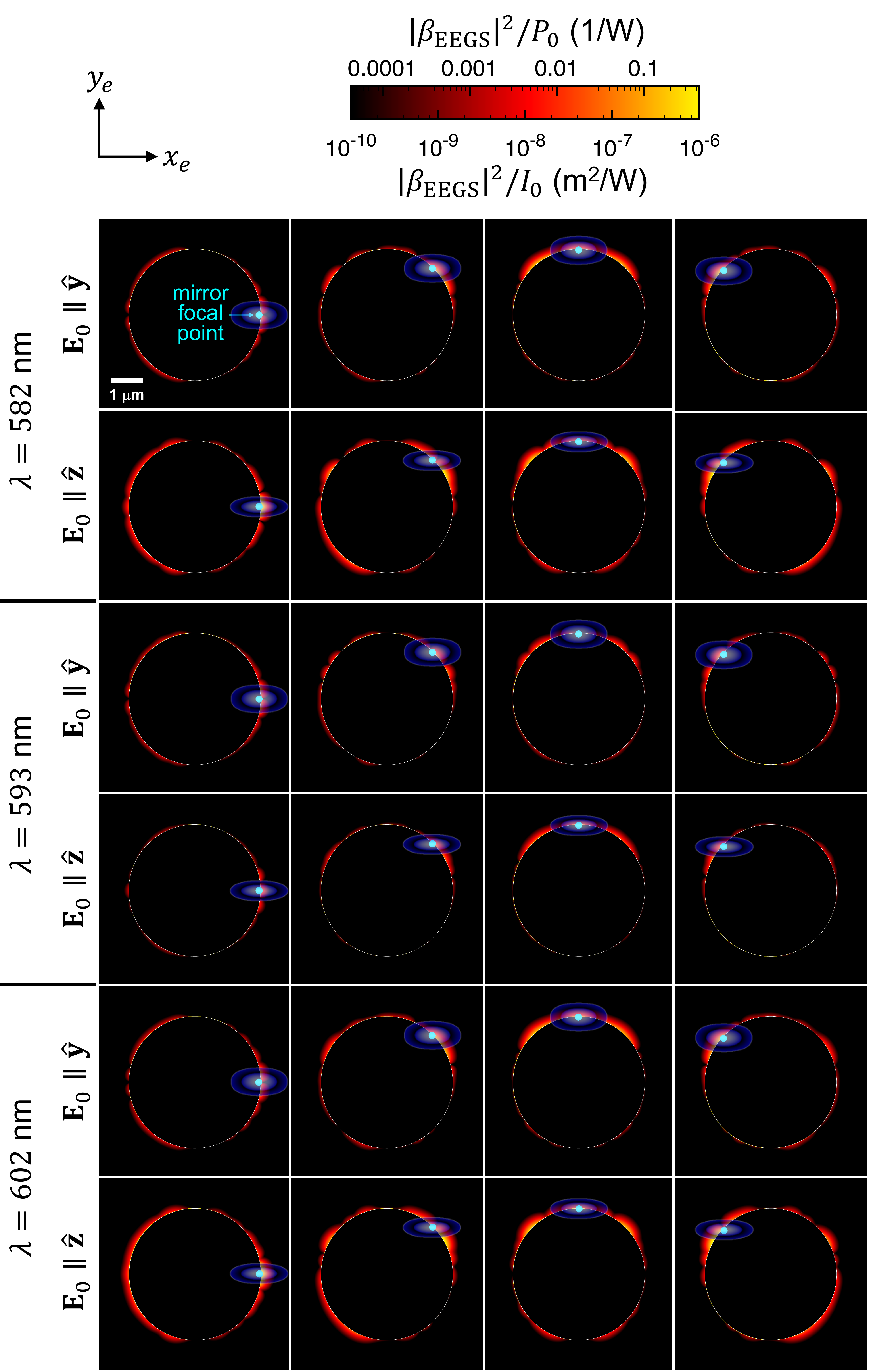}}
\caption{{\bf Same as Fig.~\ref{FigS7}, but with a smaller mirror defined by $x_0=0.5\,$mm (see Fig.~\ref{FigS2}).} The smaller mirror area ($A=2.30\,$mm$^2$ when projected on the $yz$ plane) produces larger laser focal spots. The color scale for $|\beta_{\rm EGGS}(\omega)|^2/P_0$ runs over similar values as in Fig.~\ref{FigS7} (i.e., we have a similar coupling to the electron for the same power incident on the mirror in each case), while the scale for $|\beta_{\rm EGGS}(\omega)|^2/P_0$ shows numbers that are roughly one order of magnitude smaller than in Fig.~\ref{FigS7}, in agreement with the ratio of projected mirror areas between both mirrors.}
\label{FigS8}
\end{figure*}

% Figure S9 -----------------------------------------------
\begin{figure*}
\centering{\includegraphics[width=0.7\textwidth]{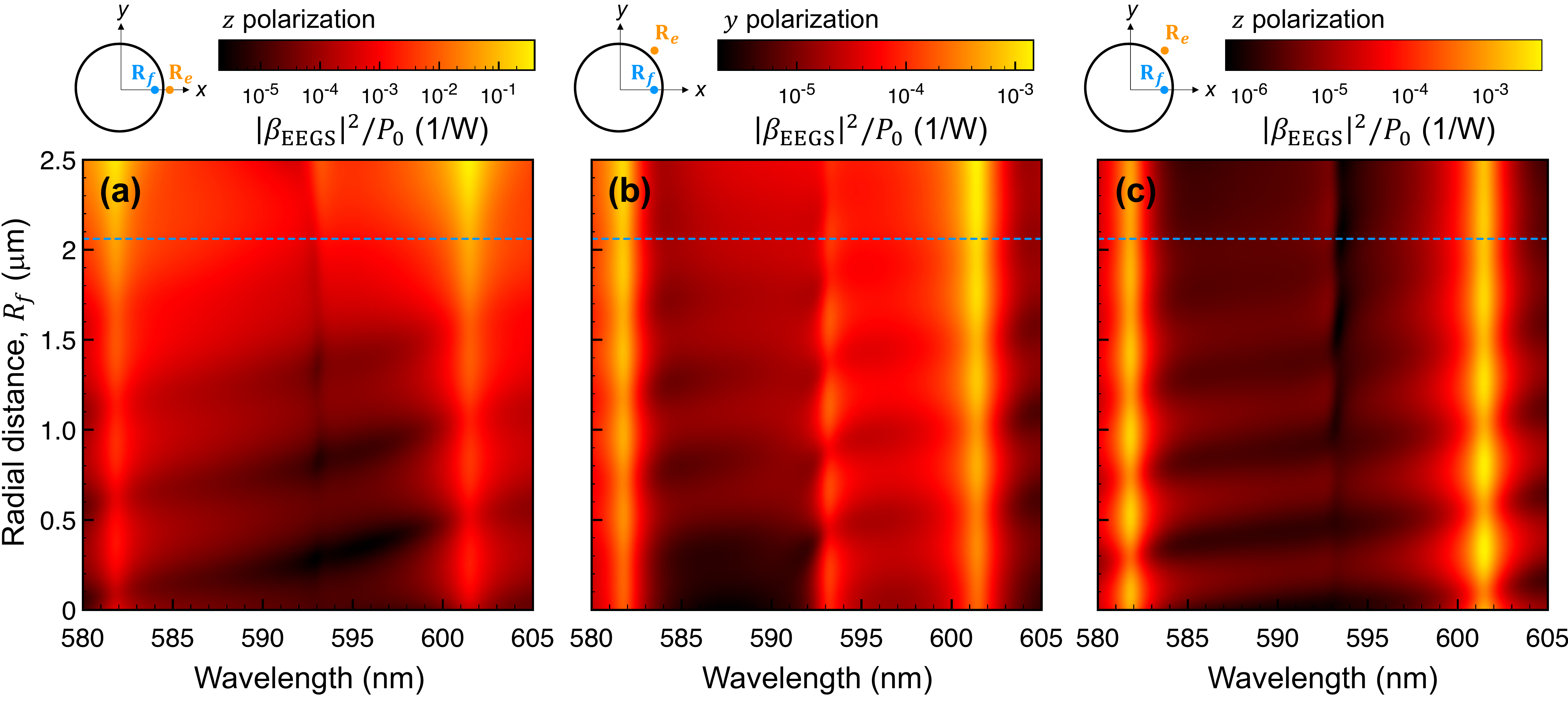}}
\caption{{\bf Dependence of the EEGS spectra on the position of the optical focal spot.} We plot the EEGS probability for the silica sphere of Fig.~\ref{FigS4} (4122\,nm diameter) as a function of laser wavelength (horizontal axes) and radial position of the laser focal spot $R_f$. The latter is placed along the positive $x$ axis. The sphere edge is indicated by dashed horizontal lines. The e-beam passes 50\,nm outside the sphere surface, crossing either the $x$ axis in (a) or a line that forms $45^\circ$ with the $x$ axis in (b,c) (see upper-left insets). The laser is polarized along $z$ in (a,c) and along $y$ in (b) (see Fig.~\ref{FigS2}). The silica permittivity is supplemented with an additional imaginary part of $10^{-3}\,\ii$ to slightly broaden the resonances.}
\label{FigS9}
\end{figure*}

% --- Table peaks -----------------------------------------
\begin{table*}[t]
\begin{tabular}{c|c|c|c|c} \hline
$\begin{matrix} \text{mode}\\ \text{wavelength} \end{matrix}$                                 \quad&\quad
$\begin{matrix} \text{4122\,nm}\\ \text{silica sphere}\\ \text{grazing}         \end{matrix}$ \quad&\quad
$\begin{matrix} \text{4122\,nm}\\ \text{silica sphere}\\ \text{50\,nm distance} \end{matrix}$ \quad&\quad
$\begin{matrix} \text{8636\,nm}\\ \text{PS sphere}\\     \text{grazing}         \end{matrix}$ \quad&\quad
$\begin{matrix} \text{8636\,nm}\\ \text{PS sphere}\\     \text{50\,nm distance} \end{matrix}$ \\
\hline
&&&&\\$582\,$nm \quad&\quad 0.0100 \quad&\quad 0.0031 \quad&\quad -      \quad\quad&\quad -      \\
&&&&\\$593\,$nm \quad&\quad 0.0032 \quad&\quad 0.0096 \quad&\quad -      \quad\quad&\quad -      \\
&&&&\\$602\,$nm \quad&\quad 0.0219 \quad&\quad 0.0070 \quad&\quad -      \quad\quad&\quad -      \\
&&&&\\$592\,$nm \quad&\quad -      \quad&\quad -      \quad&\quad $8.5\times10^{-5}$ \quad\quad&\quad $2.8\times10^{-5}$ \\
&&&&\\$594\,$nm \quad&\quad -      \quad&\quad -      \quad&\quad 0.0003 \quad\quad&\quad 0.0001 \\
&&&&\\ \hline
\end{tabular}
\caption{CL probabilities for 200\,keV electrons passing either grazingly or at a distance of 50\,nm from the surface of the 4122\,nm sphere of Fig.~\ref{FigS4} or the 8636\,nm PS sphere of Fig.~\ref{FigS5}. The probabilities are integrated over the spectral peaks at the wavelengths indicated in the left column. Incidentally, the CL probability for the PS sphere is nearly independent on the parameter $\eta$ used to introduce losses (see Fig.~\ref{FigS5}) when comparing $\eta=0$ and $\eta=10^{-4}$, although the EELS probability is substantially increased with the latter choice.}
\label{TableS1}
\end{table*}

% =========================================================
% --- Results: EEGS ---------------------------------------
% =========================================================
\section{EEGS in a large silica sphere}

We start by calculating the EELS probability for the 4122\,nm silica sphere considered in Sec.~\ref{Miemodes} (Fig.~\ref{FigS6}a), which coincides with the CL emission probability because the material is lossless. Using the elements introduced in Sec.~\ref{sphericalspecimen}, these probabilities are given by \cite{paper149}
\begin{align}
\left[ \begin{array}{c}
\Gamma_{\rm EELS}(\omega) \\ \\
\Gamma_{\rm CL}(\omega)
\end{array} \right]
=
\frac{\alpha}{\omega}\sum_{l=1}^\infty\sum_{m=-l}^l \frac{1}{l(l+1)} K_m^2\bigg(\frac{\omega R_e}{v\gamma}\bigg)
\left(
\frac{4m^2v^2}{c^2}|A_{lm}|^2
\left[\begin{array}{c}
{\rm Im}\{t_l^M\}     \\ \\
|t_l^M|^2
\end{array} \right]
+\frac{1}{\gamma^2}|B_{lm}|^2
\left[\begin{array}{c}
{\rm Im}\{t_l^E\}    \\ \\
|t_l^E|^2
\end{array} \right]
\right), \label{GammaEELS}
\end{align}
such that the total frequency-integrated probability per electron is $\int_0^\infty d\omega\,\Gamma_{\rm EELS,CL}(\omega)$. Incidentally, these expressions are made unit-independent by using the fine structure constant $\alpha\approx1/137$. The calculated spectrum is dominated by the two electric Mie modes discussed in Fig.~\ref{FigS4}c, with an additional small contribution from the intermediate magnetic mode.

Likewise, the EEGS probability $|\beta_{\rm EGGS}(\omega)|^2$, obtained from Eqs.~(\ref{fCL}) and (\ref{betaEEGSfinalb}), is dominated by the same two electric modes (Fig.~\ref{FigS6}b). In particular, for the configuration A (see inset), the magnetic mode only produces a residual feature in the EEGS spectral profile. Incidentally, the effect of light concentration by the mirror gives rise to an enhancement by five orders of magnitude in the EEGS probability relative to direct light plane-wave illumination for a given value of the light intensity $I_0$ (cf. Fig.~\ref{FigS6}b and Fig.~\ref{FigS6}c).

In Fig.~\ref{FigS7}, we present spatial maps of the calculated EEGS probability $|\beta_{\rm EGGS}(\omega)|^2$ for the 4122\,nm silica sphere. Each pixel in the map corresponds to a different e-beam position for a fixed light focal spot (blue dots). We only consider e-beam positions that do not intersect the sphere. Results are presented for the three Mie modes discussed in Fig.~\ref{FigS4}c and for four different positions of the focal spot. These plots demonstrate that substantial electron-light coupling takes place even when the e-beam passes at a large distant (but still near the sphere surface) relative to the focal spot. This is a manifestation of the delocalized nature of the Mie modes under consideration.

Incidentally, when reducing the size of the mirror, the EEGS probability $|\beta_{\rm EGGS}(\omega)|^2$ decreases roughly in proportion to the mirror area, as shown in Fig.~\ref{FigS8}. However, when normalized to the total laser power $P_0$ collected by the mirror (see upper color scale), we find a similar order of magnitude of $|\beta_{\rm EGGS}(\omega)|^2/P_0$ as for the larger mirror, indicating that the electron couples to Mie modes that are relatively delocalized, and thus, although a larger mirror is beneficial for delivering more power in practice, the size of the focal spot is not too relevant for the present specimen.

Another signature of optical field delocalization is found by examining the EEGS spectra produced at a constant e-beam position as we move the optical spot. The results shown in Fig.~\ref{FigS9} reveal two interesting trends. (1) When the e-beam and laser spot cross the same azimuthal direction (the positive $x$ axis in Fig.~\ref{FigS9}a), the EEGS probability takes comparatively large values, which tend to increase as the optical spot is placed further apart from the sphere center, such that it is able to excite Mie resonances of higher angular momentum following the rule $l\sim\omega R_f/c$, which yields an orbital number $l\sim26$ at $R_l=2.5\,\mu$m, matching well the two prominent modes at $l=26$ and 27 observed in Fig.~\ref{FigS9} (see Fig.~\ref{FigS6}). (2) In addition, when the azimuthal positions of the e-beam and optical spots are different (Fig.~\ref{FigS9}b,c), such radial dependence becomes much weaker and the EEGS probability is comparatively reduced, indicating that the e-beam misses the path followed by ballistic-like whispering gallery modes launched along a sphere meridian when the laser beam is grazingly incident.

Finally, we compare the CL emission probability integrated over frequency within the Mie resonances under consideration for the silica and PS spheres considered in Figs.~\ref{FigS4} and \ref{FigS5}. The results, presented in Table~\ref{TableS1}, show that the probability is much higher in the silica particle, thus providing an additional piece of information to explain why only this one can be resolved in our CL experiments.

% =========================================================
% --- spectral resolution ---------------------------------
% =========================================================
\section{On the question of spectral resolution and signal strength in EELS, CL, and EEGS}

The spectral resolution in EELS is generally limited by the width of the zero-loss peak and the energy analyzer. In contrast, the resolution in CL depends on the precision of the light spectrometer. Finally, the EEGS resolution is controlled by the width of the external laser. One could argue that CL and EEGS should be ultimately comparable in precision, as they rely on optical tools. However, a problem arises regarding the ability of these techniques to detect narrow resonances: indeed, our experimental CL system is unable to observe the narrow resonances of the PS particle; in contrast, EEGS can neatly resolve them with observed quality factors $Q\sim10^4$. We also note that for the silica sphere one observes $Q\sim100$, and both techniques allow us to resolve the resonances, although EEGS provides a superior spectral precision.

To address this puzzle, we argue that EEGS has the advantage that, under continuous-wave illumination, the spectral width of the laser can be made much narrower than the resonance, so that the electrons interact with a strongly enhanced field. For large quality factors $Q$ such as those of the PS sphere, the CL spectrometer is eventually unable to resolve the spectral shape. Even if an ultraprecise spectrometer is used, capable of resolving resonances with large $Q$, we find it instructive to compare in what follows the total CL signal integrated over the resonance as a way to quantify the number of electron counts in CL relative to EEGS.

For simplicity, we consider a narrow resonance of frequency $\omega_0$ and quality factor $Q$ in the specimen, for which the spectral dependence of the CL far-field amplitude ${\bf f}^{\rm CL}_{\rr}(\omega)$ in Eq.~(\ref{betaf}) can generally be approximated as a Lorentzian, such that
\begin{align}
\big|{\bf f}^{\rm CL}_{\rr}(\omega)\big|^2\approx\frac{\big|{\bf f}^{\rm CL}_{\rr}(\omega_0)\big|^2}{4Q^2(\omega/\omega_0-1)^2+1}.
\nonumber
\end{align}
Now, the angle- and frequency-dependent photon emission probability in CL is given by \cite{paper149} $\Gamma(\rr,\omega)=(c/4\pi\hbar\omega)\,\big|{\bf f}^{\rm CL}_{\rr}(\omega)\big|^2$, so that the total probability (i.e., integrated over emission angles and frequencies within the resonance peak) reduces to
\begin{align}
\Gamma_{\rm CL}=
\int d^2\rr \int_{\rm peak}\!\!\!\!\! d\omega\,\big|{\bf f}^{\rm CL}_{\rr}(\omega)\big|^2\approx \frac{c}{8Q\hbar} \int d^2\rr\,\big|{\bf f}^{\rm CL}_{\rr}(\omega_0)\big|^2.
\nonumber
\end{align}
Likewise, from the EEGS probability $\Gamma_{\rm EEGS}(\omega)=|\beta_{\rm EGGS}(\omega)|^2$, using Eq.~(\ref{betaf}) for plane-wave illumination, we find
\begin{align}
\Gamma_{\rm EEGS}=\frac{c^4}{8\pi\hbar^2\omega_0^4}\,\big|E^{\rm ext}\big|^2\int d^2\rr\,\big|{\bf f}^{\rm CL}_{\rr}(\omega_0)\big|^2,
\nonumber
\end{align}
where, for the sake of our argument, we assume resonant illumination ($\omega=\omega_0$) and introduce an average over orientations of the light field $\Eb^{\rm ext}$ and the directions $\rr$ from which light is coming (i.e., we integrate over directions $\rr$, divide by the full $4\pi$ solid angle of the sphere, and also divide by a factor of 2 to account for the average over light polarizations). We have now quantities that can directly be compared to find the dimensionless quantity
\begin{align}
\mathcal{R}^{\rm EEGS}_{\rm CL}=\frac{Qc^3\big|E^{\rm ext}\big|^2}{\pi\hbar\omega_0^4} = \frac{2Qc^2I^{\rm ext}}{\hbar\omega_0^4},
\nonumber
\end{align}
which is the ratio of EEGS-to-CL counts normalized per incident electron. Here, we use the relation $I^{\rm ext}=c\big|E^{\rm ext}\big|^2/2\pi$ between the incident light electric field amplitude and the light intensity. For $\hbar\omega_0=2\,$eV, this expression can be written as
\begin{align}
\mathcal{R}^{\rm EEGS}_{\rm CL}=Q\,\frac{I^{\rm ext}}{5\times10^{10}\,{\rm W}/{\rm m}^2}.
\nonumber
\end{align}
Then, for a plausible light intensity of $10^8\,$W/m$^2$ in our experiment, the ratio with $Q=200$ (silica sphere) is $\mathcal{R}^{\rm EEGS}_{\rm CL}=0.4$, and therefore, both CL and EEGS render comparable signals. In contrast, with $Q=10^4$ (PS sphere), the ratio becomes $\mathcal{R}^{\rm EEGS}_{\rm CL}=20$, so that EEGS is much stronger and capable of resolving the resonances, whereas the CL signal is too weak. Finally, we note that the actual ratio $\mathcal{R}^{\rm EEGS}_{\rm CL}=0.4$ should depend on the specific illumination directions, and also on the effect of the mirror.

\end{widetext} %---arxiv

% --- acknowledgments -------------------------------------
\begin{acknowledgments}
We thank T. Lovejoy for discussions on the functioning of the fast blanker on the Chromatem machine. We are also thankful to H. Lourenço-Martins, S. Meuret, and A. Arbouet for fruitful discussions. This project has been funded in part by the European Union through the Horizon 2020 Research and Innovation Program (grant agreement No. 101017720 (EBEAM)), the French National Agency for Research under the program of future investment TEMPOS-CHROMATEM (reference No. ANR-10-EQPX-50) and QUENOT (ANR-20-CE30-0033), the Spanish MICINN (PID2020–112625GB-I00 and SEV2015-0522), the ERC (Advanced Grants 789104-eNANO and 787510-4DBIOSERS), the Catalan CERCA Program, and Fundaci\'o Privada Cellex. \textbf{Author contributions:} M.K., Y.A., O.S., M.T., L.H.G.T, and F.J.G.d.A designed the experiment. Y.A., M.T., and J.-D.B. developed the light injection and synchronization setup with support from L.F.Z. and L.T. Y.A. performed experiments with support from L.H.G.T, X.L., and M.K. Y.A. analyzed the data with support from L.H.G.T. F.J.G.d.A. developed the theory and carried out numerical simulations with support from E.D. All authors participated in the results analysis and contributed to the
preparation of the manuscript.
\end{acknowledgments}

%\bibliography{refsL.bib}


\end{document}